\documentclass[aps,twocolumn,showkeys]{revtex4-2}
\usepackage[latin1,utf8]{inputenc} 
\usepackage{amssymb}
\usepackage{amsfonts} 
\usepackage{graphicx} 
\usepackage{float}
\usepackage{amsthm} 
\usepackage{amsmath} 
\usepackage{bm}
\usepackage{epsfig}
\usepackage{hyperref}
\usepackage{color}
\usepackage{bbold}
 \usepackage{bbm}
\usepackage[T1]{fontenc}
\usepackage{young}
\usepackage{youngtab}
\usepackage{xcolor}
\usepackage{braket}

\def\ic{\mathrm{i}}

\def \bc {\begin{center}}
\def \ec {\end{center}}
\def \bi {\begin{itemize}}
\def \ei {\end{itemize}}

\def \ba {\begin{array}}
\def \ea {\end{array}}

\def \bea {\begin{eqnarray}}
\def \eea {\end{eqnarray}}

\def \be {\begin{equation}}
\def \ee {\end{equation}}

\newcommand{\la}{\langle}
\newcommand{\ra}{\rangle}

\def\tr {\mathrm{tr}}

\def\bc {\bar{\beta}}

\def\nb{{\vec{n}}}
\def\zb{\bm{z}}

\def\rmu{\mathrm{U}}

\def\2cat{{\scriptstyle\mathrm{2CAT}}}
\def\3cat{{\scriptstyle\mathrm{3CAT}}}

\begin{document}

\title{Localization measures of parity adapted U($D$)-spin coherent states applied to the phase space analysis of the $D$-level Lipkin-Meshkov-Glick model}

\author{Alberto Mayorgas}
\email{albmayrey97@ugr.es}
\affiliation{Department of Applied Mathematics, University of  Granada,
	Fuentenueva s/n, 18071 Granada, Spain}
\author{Julio Guerrero}
\email{jguerrer@ujaen.es}
\affiliation{Department of Mathematics, University of Jaen, Campus Las Lagunillas s/n, 23071 Jaen, Spain}
\affiliation{Institute Carlos I of Theoretical and Computational Physics (iC1), University of  Granada,
	Fuentenueva s/n, 18071 Granada, Spain}
\author{Manuel Calixto}
\email{calixto@ugr.es}
\affiliation{Department of Applied Mathematics, University of  Granada,
	Fuentenueva s/n, 18071 Granada, Spain}
\affiliation{Institute Carlos I of Theoretical and Computational Physics (iC1), University of  Granada,
	Fuentenueva s/n, 18071 Granada, Spain}

\date{\today}

\begin{abstract}
	\vspace{1cm}
	\section*{Abstract}
%
 We study phase-space properties of critical, parity symmetric, $N$-quDit systems undergoing a quantum phase transition (QPT) in the thermodynamic $N\to\infty$ limit. The $D=3$ level (qutrit)  Lipkin-Meshkov-Glick (LMG) model is eventually examined as a particular example.  
 For this purpose, we consider U$(D)$-spin coherent states (DSCS), generalizing the standard $D=2$ atomic coherent states, to define the coherent state representation $Q_\psi$ (Husimi function) of a symmetric $N$-quDit state $|\psi\ra$ in the phase space $\mathbb CP^{D-1}$ (complex projective manifold). DSCS are good variational aproximations to the ground state of a $N$-quDit system, specially in the $N\to\infty$ limit, where the  discrete parity symmetry $\mathbb{Z}_2^{D-1}$ is spontaneously broken. For finite $N$, parity can be restored by projecting DSCS onto $2^{D-1}$ different parity invariant subspaces, which define generalized ``Schr\"odinger cat states'' reproducing quite faithfully  low-lying  Hamiltonian eigenstates obtained by numerical diagonalization. Precursors of the QPT are then visualized for finite $N$ by plotting the Husimi function of these parity projected DSCS in phase space, together with their Husimi moments and Wehrl entropy, in the neighborhood of the critical points. These are good localization measures and markers of the QPT.

\end{abstract}

%
%
%
%
%


\maketitle

\section{Introduction}

Information theoretic and statistical measures together with phase space methods  have proved to be useful in the description and characterization of quantum phase transitions (QPTs). For example, in the traditional Anderson metal-insulator transition \cite{PhysRev.109.1492,Aulbach_2004,RevModPhys.80.1355}, where Hamiltonian eigenfunctions underlie strong fluctuations. 
Phase space methods are a fundamental tool in quantum optics \cite{Schleich}, providing connections between quantum mechanics (in the so-called Wigner/Weyl/Moyal scheme \cite{FranklinSchroeck}) and classical statistical mechanics. This connection is often established through  (quasi-classical, 
minimum uncertainty) coherent states (CSs). The best known CSs are the canonical (harmonic oscillator) CSs introduced long time ago by Schr\"odinger \cite{Schrodinger1926} and later used by Glauber to study the radiation field \cite{PhysRev.131.2766}. Canonical CSs are linked to the Heisenberg-Weyl group (with the typical Lie algebra canonical commutation relations $[q,p]=\ic\hbar$) and can be seen as a group action/displacement on the vacuum. Replacing the Heisenberg-Weyl group by the rotation group SU(2) (with angular momentum commutation relations $[J_x,J_y]=\ic\hbar J_z$ and cyclic permutations), we get the so called spin-$j$, atomic or Bloch  CSs \cite{RevModPhys.62.867,GilmorePhysRevA.6.2211}. From this perspective, 
generalizations to arbitrary (finite-dimensional) Lie groups $G$ provide further families of CSs (we address the reader to the standard reference \cite{Perelomov}). In particular, this article is involved with the generalization from $\rmu(2)$ to $\rmu(D)$, which is in the heart of the generalization from qubits (physically represented by two-level/component atom/particle quantum systems) to quDits ($D$-level quantum systems).

Canonical CSs provide complex analytic (Bargmann, phase space) representations of quantum states and operators in quantum mechanics \cite{Vourdas_2006}. Among all phase-space quasi-probability distribution functions (playing a role similar to genuine probability distributions of statistical mechanics), the more popular are Wigner $W$, Husimi $Q$ and Glauber-Sudarshan $P$ (also called Berezin's covariant and contravariant symbols, respectively) functions, usually associated with the symmetric, antinormal and normal ordering of position and momentum operators, respectively \cite{Leonhardt1997,Schleich,Curtright2014}. Although Wigner function is perhaps more popular, Husimi function can be more easily extended to general phase spaces associated to coset spaces $X=G/H$ of a symmetry  Lie group $G$ for an isotropy subgroup $H\subset G$. This will be our case, with $G=\rmu(D)$ the unitary group of degree $D$, and phase space $X=U(D)/U(D-1)=\mathbb{C}P^{D-1}$ the complex projective space generalizing the Bloch sphere $\mathbb{S}^2=\mathbb{C}P^1$ for $D=2$. This case is linked to the totally symmetric (bosonic) representation of $\rmu(D)$, to which we are going to restrict ourselves here (see \cite{nuestroPRE} for other phase spaces like the flag manifold $\rmu(D)/U(1)^D$ linked to more general fermion mixtures and Young tableaux).

Given a CS system $\{|z\ra, z\in X\}$, the Husimi function of a density matrix $\rho$ is the phase space $X$ valued function $Q_\rho(z)=\la z|\rho|z\ra$. In an attempt to build bridges between classical and quantum entropies, and even though $Q_\rho(z)$ is only a semiclassical quasi-probability distribution function, a semiclassical Shannon-like entropy was defined by Wehrl  \cite{wehrl1979} as $\mathcal{S}_W(\rho)=-\int Q_\rho(z)\log Q_\rho(z)d\mu_X(z)$, with $d\mu_X(z)$ a $G$-invariant measure on the phase space $X$. Wehrl's entropy measures the area occupied by the quantum state $\rho$ in phase space; actually, moments $M_\nu$ of $Q_\rho$ (and their associated R\'enyi-Wehrl entropies \cite{Florian.PhysRevA.69.022317,Gnutzmann_2001,Giovannetti.PhysRevA.70.022328}), like the so called inverse participation ratio $M_2$, also measure the localization of $\rho$ in phase space and are easier to compute. 

For a critical quantum system described by a Hamiltonian $H(\lambda)$ depending on a control parameter $\lambda$,  abrupt changes in the Wehrl entropy of the ground state (as a function of $\lambda$) usually provide good indicators of the existence of a quantum phase transition (QPT) around a critical point $\lambda_c$, even for a finite number $N$ of particles. Moreover, Wehrl entropy can be also used to identify the order of a QPT \cite{PhysRevE.92.052106}, as an alternative definition to the standard Ehrenfest classification based on discontinuities of the derivatives of the ground state energy density with respect to $\lambda$ in the thermodynamic limit $N\to\infty$. Husimi function and its Wehrl entropy have already given a good phase space description of interesting quantum critical systems like Bose-Einstein condensates \cite{citlali}, the Dicke model of superradiance for two-level \cite{PhysRevA.85.053831,real2013} and three-level \cite{Castanos_2018} atoms, the U(3) vibron model of molecular benders \cite{PhysRevA.86.032508}, the U(4) bilayer quantum Hall system \cite{Calixto_2018}, the U(2) (two-level) ubiquitous Lipkin-Meshkov-Glick (LMG) model \cite{Romera_2014,Romera_2017,PhysRevB.74.104118}, etc. Here we want to extend the scope of applicability of these phase space methods to symmetric multi-quDit systems (like $D$-level atom models) described by a $\rmu(D)$ invariant LMG model. In addition to the obvious technical complication, $\rmu(D)$ provides some novelties and a much richer structure 
that is not possible to grasp starting from $\rmu(2)$. In particular, the standard discrete parity symmetry group $\mathbb{Z}_2=\{0,1\}$, which is spontaneously broken in the thermodynamic limit for second order QPTs of $D=2$ level systems, now becomes $\mathbb{Z}_2^{D-1}$ and provides more case studies of Schr\"odinger cat states than the standard even and odd ones of the literature \cite{Dodonovcat,DODONOV1974597,Manko-MultimodeCats1995,TwoModeSU2-SU11-CAT}, in the sense of quantum superpositions of weakly overlapping quasiclassical (coherent) states, the most symmetric one mimicking the structure of the ground state in the highly interacting quantum phase (see later in Section \ref{LMGsec} and \cite{L_pez_Pe_a_2015} for previous studies on Dicke models of three-level atoms interacting with one-mode radiation field).

The organization of the article is as follows. In Sec. \ref{LMGsec1} we introduce the $D$-level LMG model and particularize it for the cases $D=2$ (qubits) and $D=3$ (qutrits). A brief discussion about the Fock basis and the discrete parity symmetry $\mathbb{Z}_2^{D-1}$ is also included. In Sec. \ref{statesymmatsec} we define U($D$)-spin coherent states $|\bm{z}\ra$ (DSCSs for brevity) labelled by points $\bm{z}\in\mathbb CP^{D-1}$ in phase space; we also compute the  DSCS matrix elements $\la \bm{z}|S_{ij}|\bm{z}'\ra$ of $\rmu(D)$-spin operators $S_{ij}, i,j=1,\dots,D$, and we project DSCSs $|\bm{z}\ra$ into the $2^{D-1}$ invariant subspaces $\mathbbm{c}$ of the parity symmetry group $\mathbb{Z}_2^{D-1}$, introducing the notion of ``$\mathbbm{c}$-parity $\rmu(D)$ Schr\"odinger cat states'' $|\bm{z}\ra_\mathbbm{c}$ (called $\mathbbm{c}$-DCAT states, for short). Then, in Sec. \ref{Husimisec}, the traditional Husimi function $Q_\psi(z)=|\la z|\psi\ra|^2$  of a quantum state $|\psi\ra$ in the standard phase space $\mathbb{C}\ni z$ (for canonical, harmonic oscillator or  Heisenberg-Weyl coherent states) is extended to the phase space $\mathbb CP^{D-1}\ni\bm{z}$ using DSCSs $|\bm{z}\ra$ and a convenient Harr integration measure, which allows to define $\nu$-moments of the Husimi function and the Wehrl entropy as usefull localization measures in phase space. These measures are computed in the case of DSCS and  $\mathbbm{c}$-DCAT states, including their thermodynamic limit $N\to\infty$. The Appendices \ref{app1} and \ref{app2} show in more detail some of the long calculations of this section. In Sec. \ref{LMGsec} we focus on the $D=3$ level LMG Hamiltonian for symmetric qutrits and the minimization of its energy surface in the limit $N\to\infty$ using DSCSs as variational states. The degeneration of the ground state in the thermodynamic limit and the QPTs make their apparition here, but are not discussed in depth until the next two sections. In Sec. \ref{Sec:Fidelity}, the variational ground state obtained in the previous section is projected on parity $\mathbbm{c}$ subspaces and the corresponding $\mathbbm{c}$-DCATs are compared to the  low-lying Hamiltonian eigenstates of the LMG model obtained by numerical diagonalization for finite $N$. This procedure  (projection after energy minimization) provides a fairly good variational aproximation to the ground state in terms of the completely even, $\mathbbm{c}=\mathbb{0}$, DCAT state, but not so precise for first excited states in terms of DCAT states of other parities $\mathbbm{c}$, for which we try a proper overlap maximization (fidelity) procedure. 
 In Sec. \ref{Sec:Localiz_measures}, the Husimi function and the localization measures of the Sec. \ref{Husimisec} are employed to visualize  how the variational and the numerical eigenstates split into Gaussian-like wave packets throughout the three different quantum phases of the $D=3$ level LMG model. The Inverse Participation Ratio (Husimi second moment) and the Wehrl entropy are also used to quantify the overlap of these packets, and hence the localization/spread  of the low-lying Hamiltonian eigenstates in phase space  is compared  to that of  DSCS and $\mathbbm{c}$-DCAT variational states. Finally, in Sec. \ref{conclusec} we present the main conclusions of this work.

\section{$D$-level LMG model Hamiltonian and parity symmetry}\label{LMGsec1}


The original ($D=2$ levels/modes) LMG schematic shell model appeared in nuclear physics  \cite{lipkin1,lipkin2,ring} 
to describe the quantum phase transition from spherical to deformed shapes in nuclei. Since then, it is an ubiquitous model that appears in a multitude of physical contexts.
For example, the Hamiltonian of an anisotropic XY
Ising model, with $\mu=1,\dots,N$ lattice sites, in an external transverse magnetic field $\varepsilon$ with infinite-range constant
interactions
\begin{equation}
H_{XY}=  \varepsilon \sum_{\mu=1}^N \sigma_z^{(\mu)}
+  \sum_{\mu<\nu} \lambda_x \sigma_x^{(\mu)} \sigma_x^{(\nu)}  +  \sum_{\mu<\nu} \lambda_y\sigma_y^{(\mu)} \sigma_y^{(\nu)}\:,
\label{Hlmgeneralpauli}
\end{equation}
[$\sigma_{x,y,z}^{(\mu)}$ denote the Pauli matrices at site $\mu$] adopts the form of the two-level LMG schematic shell model Hamiltonian  \cite{lipkin1,lipkin2}
\begin{equation}
H_2 = \varepsilon J_z+\frac{\lambda_1}{2}(J_+^2+J_-^2)+\frac{\lambda_2}{2}(J_+J_-+J_-J_+)\:
\label{hamU2}
\end{equation}
when written in terms of the  SU$(2)$ angular momentum collective operators 
\be
\vec{J}=(J_x,J_y,J_z)= \sum_{\mu=1}^N (\sigma_x^{(\mu)},\sigma_y^{(\mu)},\sigma_z^{(\mu)}),
\ee
and $J_\pm=(J_x\pm\ic J_y)/2$, as usual. We could also think of a model describing 
a system of $N$ interacting two-level identical atoms (symmetric ``qubits''). 
Long-range constant interactions make this Hamiltonian translation invariant, that is, it is 
symmetric under permutation of lattice sites $\mu\leftrightarrow\nu$ (or permutation of atoms/qubits). Therefore, 
the Hamiltonian does not couple different angular momentum sectors $j=N/2,N/2-1,\dots, 1/2$ or $0$ (for odd or even $N$, respectively) and it is a common practice 
to restrict oneself to the largest (fully symmetric) sector $j=N/2$ to which the ground state of the system belongs. This restriction 
reduces the size of the Hamiltonian matrix to be diagonalized from $2^N$ to $N+1=2j+1$ and assumes that $D=2$-level atoms/qubits are indistinguishable. 
For this case, it is convenient to use a Jordan-Schwinger realization of angular momentum operators in terms of bilinear products of bosonic creation $a^\dag_i$ and annihilation $a_j$ operators as
\be
S_{ij}=a^\dag_i a_j, \; i,j=0,\dots,D-1,\label{UDgen}
\ee
where we are already extending to arbitrary $D$-level atom systems with $\rmu(D)$ symmetry. For example, for $D=2$ we recover 
$J_+=S_{10}, J_-=S_{01}$, $J_z=\frac{1}{2}(S_{11}-S_{00})$ and the conserved total number $N$ of particles $C_1=S_{00}+S_{11}$ [the linear Casimir operator of $\rmu(2)$]. $\rmu(D)$-spin operators $S_{ij}$ fulfill the commutation relations 
\be
\left[S_{{ij}},S_{{kl}}\right]=\delta_{{jk}} S_{{il}} -\delta_{{il}} S_{{kj}}.\label{commurel}
\ee
The LMG Hamiltonian $H_2$ in \eqref{hamU2} for $D=2$ level systems is generalized to arbitrary $D$ levels as 
\begin{equation}
H_D = \sum_{i=0}^{D-1}\varepsilon_i(S_{i+1,i+1}-S_{ii})+\sum_{i\not=j=0}^{D-1} (\lambda_1 S_{ij}^2+\lambda_2S_{ij}S_{ji}),\label{hamUD}
\end{equation}
where $\varepsilon_i$ now denotes the energy gap between levels $i$ and $i+1$. 
The $\lambda_1$ interaction term annihilates pairs of particles in one level and creates pairs in other level, whereas the $\lambda_2$
term scatters one particle from $i\to j$ while another is scattered back from $j\to i$. The total number of particles $N=\sum_{i=0}^{D-1} S_{ii}$ (the linear Casimir operator of $\rmu(D)$) is conserved. For the sake of simplicity, we shall consider $\lambda_2=0$ 
and $\varepsilon_i=\varepsilon$ (same energy spacing between levels). Since we are interested in the thermodynamic limit $N\to\infty$, we shall also renormalize one-body interactions $\varepsilon\to \epsilon/N$ by the total number $N$ of particles, and two-body interactions $\lambda_2\to-\lambda/[N(N-1)]$ by the total number $N(N-1)$ of pairs, so that the final Hamiltonian density for us becomes
\begin{equation}
H=\frac{\epsilon}{N}(S_{D-1,D-1}-S_{00})-\frac{\lambda}{N(N-1)}\sum_{i\not=j=0}^{D-1} S_{ij}^2.\label{hamUDbis}
\end{equation}
We shall measure energy in $\epsilon>0$ units, and discuss the 
energy spectrum and the phase diagram in terms of the control parameter $\lambda$ (see later in Section \ref{LMGsec}). There are already some studies in the literature of this Hamiltonian for $D=3$ level atoms and its chaotic behavior (see e.g. 
\cite{Meredith,Lopez-Arias1989,Kus,KusLipkin,Casati,Saraceno,nuestroPRE}).

We shall consider indistinguishable atoms, so that the Hilbert space dimension reduces from $D^N$ to $\tbinom{N+D-1}{D-1}$, the dimension of the fully symmetric irreducible representation of $\rmu(D)$ (which 
coincides with the total number of compositions of $N$ into $D$ non-negative integers when order does not matter). This restriction  considerably reduces the computational complexity for large number of particles $N$ (see \cite{nuestroPRE} for the role played by other mixed permutation symmetry sectors in the 
thermodynamic limit $N\to\infty$). 
Therefore, the Hilbert space is spanned by the Bose-Einstein-Fock basis  states ($|\vec{0}\ra$ denotes the Fock vacuum)
\be
|\vec{n}\ra=|n_0,\dots, n_{D-1}\ra=
\frac{(a_0^\dag)^{n_0}\dots(a_{D-1}^\dag)^{n_{D-1}}}{\sqrt{n_0!\dots n_{D-1}!}}|\vec{0}\ra, \label{symmetricbasis}
\ee
where $n_i$ denotes the occupancy number of level $i$ (the eigenvalue of $S_{ii}$), with the restriction  $n_0+\dots+n_{D-1}=N$ (the total number of atoms/quDits). In the low-interaction regime $\lambda\ll 1$, the ground state of \eqref{hamUDbis} 
is a Bose-Einstein condensate $\tfrac{1}{\sqrt{N!}}(a_0^\dag)^{N}|\vec{0}\ra$ of $N$ atoms in the $i=0$ level, which we shall take as a reference level from now on.

These Fock states are the natural generalization of angular momentum $j=N/2$ Dicke states $|j,m\ra$ with  angular momentum third component $m=-j,\dots, j$; more explicitly
\be
|j,m\ra=|n_0=j+m,n_1=j-m\ra,
\ee
so that $m=(n_0-n_1)/2$ (the eigenvalue of $J_z=\frac{1}{2}(S_{11}-S_{00})$) represents the population imbalance between levels $i=0$ and $i=1$. The expansion of a general symmetric $N$-quDit state $\psi$ in the Fock basis will be written as
\be
|\psi\ra=\sum_{\|\vec{n}\|_1=N}\,c_{\vec{n}}|\vec{n}\ra,\label{psisym}
\ee
where the sum is restricted to $\|\vec{n}\|_1=n_0+\dots+n_{D-1}=N$. 
Collective $\rmu(D)$-spin operators  \eqref{UDgen} matrix elements in the Fock basis are easily computed as
\begin{align}
&\la\vec{m}|S_{ii}|\vec{n}\ra=n_i\delta_{\vec{m},\vec{n}}\,,\label{Sijmatrix}\\ 
	&\la\vec{m}|S_{ij}|\vec{n}\ra=\sqrt{(n_i+1)n_j}\delta_{m_i,n_{i}+1}\delta_{m_j,n_{j}-1}\prod_{k\not=i\not=j}\delta_{m_k,n_k}\,.\nonumber
\end{align}

At this point, we would like to highlight the existence of an interesting parity symmetry. Indeed, 
this symmetry of the Hamiltonian has to do with the fact that the interaction only scatters pairs of particles, thus conserving 
the parity $\Pi_j=\exp(\ic \pi S_{jj})$, even (+) or odd ($-$),  of the population $S_{jj}$ in each level $j=0,\dots,D-1$. 
Note that $\Pi_j|\vec{n}\ra=(-1)^{n_j}|\vec{n}\ra$, and therefore we have the constraint 
$\Pi_0\dots\Pi_{D-1}|\vec{n}\ra=(-1)^N|\vec{n}\ra$ which allows to write for example  $\Pi_0=(-1)^N\Pi_1\dots\Pi_{D-1}$. Hence, this discrete parity symmetry corresponds to the finite group $\mathbb{Z}_2^{D-1}=\mathbb{Z}_2\times\stackrel{D-1}{\dots}\times\mathbb{Z}_2$, with $\mathbb{Z}_2=\{0,1\}$ the usual parity group (the cyclic group of order 2). Consequently, energy eigenstates have well defined parity under $\mathbb{Z}_2^{D-1}$. 
We will see later in Sec. \ref{LMGsec} that low-lying Hamiltonian eigenstates with different parities collapse in the thermodynamic $N\to\infty$ limit, giving rise to a degenerate ground state as a consequence of a spontaneous breakdown of the parity symmetry $\mathbb{Z}_2^{D-1}$. 

Let us denote by the binary string $\mathbbm{b}=[b_1,\dots,b_{D-1}]\in\{0,1\}^{D-1}$ one of the $2^{D-1}$ elements of the parity group $\mathbb{Z}_2^{D-1}$. There are $2^{D-1}$ parity invariant subspaces labeled by  the inequivalent group characters $\mathbbm{c}=[c_1,\dots,c_{D-1}]\in\{0,1\}^{D-1}$ of the Pontryagin dual group $\widehat{\mathbb{Z}_2^{D-1}}\sim \mathbb{Z}_2^{D-1}$.
The projectors onto these invariant subspaces of definite parity $\mathbbm{c}$ are given by
\be
\Pi_\mathbbm{c}=2^{1-D}\sum_{\mathbbm{b}\in\{0,1\}^{D-1}} (-1)^{\mathbbm{c}\cdot \mathbbm{b}}\Pi^{\mathbbm{b}} \,,\label{projpar}
\ee
with $\mathbbm{c}\cdot \mathbbm{b}=c_1b_1+\dots+c_{D-1}b_{D-1}$ and 
\be 
\Pi^{\mathbbm{b}}\equiv \Pi_1^{b_1}\dots \Pi_{D-1}^{b_{D-1}}.\label{pibe}
\ee
Note that 
\be
\sum_{\mathbbm{c}\in\{0,1\}^{D-1}}\Pi_\mathbbm{c} = I\,,
\ee
the identity $I$ in the representation space. For example, for $D=2$ we have just $\Pi_{\mathbb{0}}=\Pi_\mathrm{even}$ and $\Pi_{\mathbbm{1}}=\Pi_\mathrm{odd}$ the standard projectors on even and odd parities, with 
$I=\Pi_\mathrm{even}+\Pi_\mathrm{odd}$. For general $D$, we sometimes  shall single out the totally even $\mathbb{0}=[0,\dots,0]$ and totally odd $\mathbbm{1}=[1,\dots,1]$ parity representations.

\section{$\rmu(D)$-spin coherent states and adaptation to parity}\label{statesymmatsec}

\subsection{$\rmu(D)$-spin coherent states}

$\rmu(D)$-spin coherent states (DSCSs for brevity) are defined as a generalization of standard binominal (two-mode) $\rmu(2)$-spin coherent states to the multinomial ($D$-mode) case as
\begin{equation}
	|\zb\ra^{(N)}=\frac{1}{\sqrt{N!}}\left(
	\frac{a_0^\dag+z_1a_1^\dag+\cdots+z_{D-1} a_{D-1}^\dag}{\sqrt{1+|z_1|^2+\cdots+|z_{D-1}|^2}}\right)^{N}|\vec{0}\ra,\label{cohD}
\end{equation}
so that they are labeled by $D-1$ complex numbers $z_j\in\mathbb{C}$ arranged in the column vector $\zb=(z_1,z_2,\dots,z_{D-1})^t\in \mathbb{C}^{D-1}$. Properly speaking, 
this really corresponds to a certain patch of the complex projective manifold $\mathbb CP^{D-1}$, 
which results when choosing $i=0$ as a reference level; see e.g. \cite{QIP-2021-Entanglement} for more information about other choices and patches. DSCSs are also labeled by the total number of particles $N$ [also labelling a specific symmetric representation of U($D$)], which will be omitted as superscript in eq.\eqref{cohD} to simplify the notation, i.e. $|\zb\ra\equiv|\zb\ra^{(N)}$.

DSCSs  $|\zb\ra$ have the form of a  Bose-Einstein condensate of $D$ modes, generalizing the spin $\rmu(2)$ (binomial) coherent states of two modes introduced by \cite{Radcliffe} and \cite{GilmorePhysRevA.6.2211} long time ago.  
If we take $i=0$ as a reference energy level, then the state $|\zb=\bm{0}\ra$ would be the ground state, whereas general $|\zb\ra$ could be seen as coherent excitations. The coefficients $c_\nb(\zb)$ of the expansion \eqref{psisym} 
of $|\psi\ra=|\zb\ra$ in the Fock basis are simply 
\be
c_\nb(\zb)=\sqrt{\frac{N!}{\prod_{i=0}^{D-1} n_i!}}\frac{\prod_{i=1}^{D-1} z_i^{n_i}}{(1+\zb^\dag\zb)^{N/2}},\label{coefCS}
\ee
where $\zb^\dag\zb=|z_1|^2+\dots+|z_{D-1}|^2$ denotes the standard scalar product in $\mathbb{C}^{D-1}$. 

In general, DSCSs are not orthogonal since the scalar product 
\be \la \zb|\zb'\ra=\frac{(1+\zb^\dag \zb')^N}{(1+\zb^\dag \zb)^{N/2}(1+\zb'^\dag \zb')^{N/2}}\label{scprod}
\ee
is not necessarily zero.  However, they are a overcomplete set of states closing a resolution of the identity 
\begin{align}\label{resounity}
	1=&\,\int_{\mathbb{C}^{D-1}}|\zb\ra\la\zb|d\mu(\zb),\\
	d\mu(\zb)=&\,\frac{(D-1)!}{\pi^{D-1}}\binom{N+D-1}{N}\frac{d^2z_1\dots d^2z_{D-1}}{(1+\zb^\dag \zb)^{D}}\,,\nonumber
\end{align}
with $d^2z_i=d\Re(z_i) d\Im(z_i)$ the Lebesgue measure on $\mathbb{C}$ and $d\mu(\zb)$ the Fubini-Study measure \cite{Florian.PhysRevA.69.022317, bengtsson_zyczkowski_2006} in the corresponding complex projective space. 
This closure relation of DSCSs will be important when discussing phase space constructions.

\subsection{Coherent state operator matrix elements}

DSCS matrix elements of $D$-spin operators $S_{ij}$ are easily computed from \eqref{Sijmatrix} and \eqref{coefCS} and they are simply 
\be\label{CSEV}
\la \zb'|S_{ij}|\zb\ra=N \bar{z}'_i z_j\frac{(1+\zb'^\dag\zb)^{N-1}}{(1+\zb'^\dag\zb')^{N/2}(1+\zb^\dag\zb)^{N/2}},
\ee
where we understand $z_0=1=z'_0$. From here, DSCS matrix elements of quadratic powers of $D$-spin operators can be concisely written as

\begin{align}\label{CSEV2}
	\langle \zb'|S_{ij}S_{kl}|\zb\rangle=&\,\delta_{jk}\la \zb'|S_{il}|\zb\ra\\
	&+\frac{N-1}{N}\frac{ \la\zb'|S_{ij}|\zb\ra\la \zb'|S_{kl}|\zb\ra}{\la \zb'|\zb\ra}\,.\nonumber
\end{align}
Note that
\begin{equation}
\lim_{N\to\infty}\frac{ \langle \zb|S_{ij}S_{kl}|\zb\rangle}{\langle \zb|S_{ij}|\zb\rangle\langle \zb|S_{kl}|\zb\rangle}= 1,\label{nofluct}
\end{equation}
which means that quantum fluctuations are negligible  in the thermodynamic (classical) limit $N\to\infty$. We shall use these ingredients when computing energy surfaces in Section \ref{LMGsec}.

\subsection{Parity adapted $\rmu(D)$-spin coherent states}

DSCSs are sometimes called ``quasi-classical'' states. As we shall 
see in Section  \ref{LMGsec}, $|\zb\ra$ turns out to be a good variational state, which reproduces the energy and wave function of the ground state of multilevel LMG atom models in the thermodynamic (classical) 
limit $N\to\infty$. However, DSCSs do not display the parity 
symmetry $\mathbb{Z}_2^{D-1}$ of the LMG Hamiltonian, which is commented at the end of Section \ref{LMGsec1}. This parity symmetry is spontaneously broken  in the thermodynamic limit $N\to\infty$ due to the degeneration of the different parity states, but it should 
be restored for finite $N$  to properly reproduce the ground (and excited) state wave function properties. A parity adaptation of DSCSs can be done by applying  projectors $\Pi_{\mathbbm{c}}$ in \eqref{projpar} on 
invariant subspaces of definite parity $\mathbbm{c}$. The effect of level $i$ population parity operations $\Pi_i=\exp(\ic \pi S_{ii})$ on DSCSs reduces to 
\begin{align}\label{parityCS}
	\Pi_i|\zb\ra=|(z_1,\dots,-z_i,\dots,z_{D-1})\ra.
\end{align}
That is, $\Pi_i$ just changes the sign of $z_i$ in $|\zb\ra$. Let us denote by
\be\label{z_b}
|\zb\ra^\mathbbm{b}= \Pi^\mathbbm{b}|\zb\ra=|((-1)^{b_1}z_1,\dots,(-1)^{b_{D-1}}z_{D-1})\ra \equiv |\zb^{\mathbbm{b}}\ra,
\ee
with $\Pi^\mathbbm{b}$ in \eqref{pibe}, and by 
\be\label{DCAT}
|\zb\ra_\mathbbm{c}\equiv \frac{\Pi_{\mathbbm{c}}|\zb\ra}{\mathcal{N}(\zb)_\mathbbm{c}}=\frac{2^{1-D}}{\mathcal{N}(\zb)_\mathbbm{c}}\sum_{\mathbbm{b}\in\{0,1\}^{D-1}}(-1)^{\mathbbm{c}\cdot \mathbbm{b}}|\zb\ra^\mathbbm{b},
\ee
with $\Pi_\mathbbm{c}$ in \eqref{projpar}, the normalized projection of $|\zb\ra$ onto the parity $\mathbbm{c}$ invariant subspace, with squared normalization factor
\be
\mathcal{N}(\zb)_\mathbbm{c}^2=2^{1-D} \frac{ \sum_{\mathbbm{b}} (-1)^{\mathbbm{c}\cdot\mathbbm{b}} (1+\zb^\dag\zb^\mathbbm{b})^N}{(1+\zb^\dag\zb)^N}.\label{normcat}
\ee
We will write $|\zb\ra^{\mathbbm{b}}=|\zb^{\mathbbm{b}}\ra$ indistinctly, with $\zb^{\mathbbm{b}}=((-1)^{b_1}z_1,\dots,(-1)^{b_{D-1}}z_{D-1})$ as defined in the eq.\eqref{z_b}. The same as $\Pi^{\mathbbm{b}}$ and $\Pi_{\mathbbm{c}}$ denote different operators, do not confuse 
 $|\zb\ra^\mathbbm{b}$ with $|\zb\ra_\mathbbm{c}$, which can be seen as the dual Fourier (Walsh-Hadamard) transformed version of $|\zb\ra^\mathbbm{b}$ with 
\be
\chi_\mathbbm{c}(\mathbbm{b})=(-1)^{\mathbbm{c}\cdot \mathbbm{b}}=(-1)^{c_1b_1+\dots c_{D-1}b_{D-1}}\label{charactersz2}
\ee
the characters of the parity group $\mathbb{Z}_2^{D-1}$. The factors $(-1)^{c_ib_i}$ are the analogue of the traditional discrete Fourier transform characters $\chi_\omega(t)=e^{\ic\omega t}, \omega,t=0,\dots,M-1$ but for the 
additive group $\mathbb{Z}_M$ of integers modulo $M$ (or the multiplicative group of $M$-th roots of unity), with $M=2$ in our case. The characters \eqref{charactersz2} have some useful properties such as
\bea
	\sum_{\mathbbm{c}\in\{0,1\}^{D-1}}\chi_\mathbbm{c}(\mathbbm{b})&=&2^{D-1}\delta_{\mathbbm{c},\mathbb{0}}\,,\label{CharPropDelta}\\
	\chi_\mathbbm{c}(\mathbbm{b})&=&\chi_\mathbbm{b}(\mathbbm{c})\,,\\
	\chi_\mathbbm{c}(\mathbb{0})&=&1\,,\label{ChiProp1}\\
	\chi_\mathbbm{c}(\mathbbm{b})\chi_{\mathbbm{c}'}(\mathbbm{b})&=&\chi_{\mathbbm{c}+\mathbbm{c}'}(\mathbbm{b})\,.
	\label{CharProp}
\eea

The coefficients $c_\nb(\zb)_\mathbbm{c}$ of the expansion \eqref{psisym} 
of $|\psi\ra=|\zb\ra_\mathbbm{c}$ in the Fock basis can be derived from \eqref{DCAT} and \eqref{coefCS},
\begin{align}
	c_\nb(\zb)_\mathbbm{c}=&\,\frac{2^{1-D}}{\mathcal{N}(\zb)_\mathbbm{c}}\sum_{\mathbbm{b}\in\{0,1\}^{D-1}}(-1)^{(\mathbbm{c}+\mathbbm{n})\cdot \mathbbm{b}}c_\nb(\zb)\nonumber\\ 
	=&\,\frac{1}{\mathcal{N}(\zb)_\mathbbm{c}}c_\nb(\zb)\delta_{\mathbbm{n},\mathbbm{c}}\,,\label{coefDCAT}
\end{align}
where $\mathbbm{n}=[\text{mod}(n_1,2),\ldots,\text{mod}(n_{D-1},2)]$ is retrieved from $\nb$ removing $n_0$  and expressing it in modulo 2, and $\delta_{\mathbbm{n},\mathbbm{c}}=\delta_{\text{mod}(n_1,2),c_1}\cdots\delta_{\text{mod}(n_{D-1},2),c_{D-1}}$ is the product of Kronecker deltas.

For $D=2$, the parity adaptations $|\zb\ra_{[0]}=|\zb\ra_{+}$ and 
 $|\zb\ra_{[1]}=|\zb\ra_{-}$ of a $\rmu(2)$-spin coherent state $|\zb\ra$ (for  $\zb=(z_1)=z$) adopt the form 
\be
|z\ra_\pm=\frac{|z\ra\pm|-z\ra}{\sqrt{2\pm2\left(\frac{1-|z|^2}{1+|z|^2}\right)^N}},\label{eoCS}
\ee
and are  sometimes called even (+) and odd ($-$) ``Schr\"odinger cat states'',  since they are a quantum superposition of weakly-overlapping  (or distinguishable, i.e. $\la z|-z\ra\xrightarrow{N\to\infty}0$ for $z\not=0$)  
quasi-classical (minimal uncertainty) coherent wave packets. Hence, we shall name $\mathbbm{c}$-DCATs the $\mathbbm{c}$-parity adapted DSCSs $|\zb\ra_\mathbbm{c}$ in \eqref{DCAT} from now on.

Likewise, for $D=3$ we have $2^{D-1}=4$ parity sectors, 
\be
\mathbbm{c}=[c_1,c_2]\in\big\{[0,0], \, [0,1], \, [1,0], \, [1,1]\big\},
\ee
and therefore four Schr\"odinger cat states associated to the DSCS $|\zb\ra=|(z_1,z_2)\ra$ adopting the explicit form
\begin{align}\label{S3C}
|\zb\ra_{\mathbbm{c}}=\frac{1}{4\mathcal{N}(\zb)_\mathbbm{c}}\Big[&\,|(z_1,z_2)\ra\\
	+&\,(-1)^{c_1}|(-z_1,z_2)\ra+(-1)^{c_2}|(z_1,-z_2)\ra\nonumber\\
	+&\,(-1)^{c_1+c_2}|(-z_1,-z_2)\ra\Big]\nonumber\,,
\end{align}
with squared norm 
\begin{align}\label{S3CN}
	\mathcal{N}(\zb)_{\mathbbm{c}}^2=&\,\frac{1}{4 (1+|z_1|^2+|z_2|^2)^N}\Big[(1+|z_1|^2+|z_2|^2)^N\nonumber\\
	&\,+(-1)^{c_1} (1-|z_1|^2+|z_2|^2)^N\nonumber\\
	&\,+(-1)^{c_2} (1+|z_1|^2-|z_2|^2)^N\nonumber\\
	&\,+(-1)^{c_1+c_2} (1-|z_1|^2-|z_2|^2)^N\Big]\,.
\end{align}

Note that there are at most  $2^{D-1}$ Schr\"odinger cat states $|\zb\ra_{\mathbbm{c}}$ associated to a DSCS $|\zb\ra$ for arbitrary $\zb$. However, we can have  $\Pi_{\mathbbm{c}}|\zb\ra=0$ and $\mathcal{N}(\zb)_\mathbbm{c}=0$ when ${c}_i=1$ and $z_i=0$, so that the $\mathbbm{c}$-DCAT in \eqref{DCAT} contains an indeterminate form of type ``$0/0$''.
For instance, in the previous example with $D=2$, the odd 2CAT state becomes
\begin{align}
	\lim\limits_{z\to 0}|z\ra_-=&\,\lim\limits_{z\to 0}\frac{|z\ra\pm|-z\ra}{\sqrt{2\pm2\left(\frac{1-|z|^2}{1+|z|^2}\right)^N}}\nonumber\\
	=&\,\lim\limits_{z\to 0}\frac{\left(\frac{2\sqrt{N}}{\sqrt{(N-1)!}}z(a_0^{\dagger})^{N-1}a_1^{\dagger}+O(z^2)\right)|\vec{0}\ra}{2\sqrt{N}z+O(z^2)}\nonumber\\
	=&\,\,|{\scriptstyle n_0=N-1,\,n_1=1}\ra .\label{2CAT_limits}
\end{align}
The result is then a Fock basis state \eqref{symmetricbasis}, which codifies the antisymmetry of the odd 2CAT $|z\ra_-$ by filling the  level $i=1$ with $n_1=1$ particle. This ``transmutation'' of $\mathbbm{c}$-DCATs into Fock states for some zero components of $\bm{z}$ will be visualized  when plotting the Husimi function of the $\mathbbm{c}$-DCATs in the next section. On the other hand, the even 2CAT also transmutes  to another Fock basis state in the limit $\lim\limits_{z\to 0}|z\ra_+=|{\scriptstyle n_0=N,n_1=0}\ra $. 

It is also relevant to calculate the $z_i\to 0$ limits in the particular case of the  $\mathbbm{c}$-3CATs, as they will be used to study the variational aproach to the Hamiltonian eigenstates of the LMG U(3) model in the different quantum phases in Sec. \ref{Sec:Fidelity}. For $D=3$, the 3CAT state \eqref{S3C} has the following limits
\begin{align}
	\lim\limits_{z_1\to 0}|\zb\ra_{\mathbbm{c}}^{(N)}=&\, (a_1^{\dagger})^{c_1}|(0,z_2)\ra_{[c_2]}^{(N-c_1)}\,,\nonumber\\
	\lim\limits_{z_2\to 0}|\zb\ra_{\mathbbm{c}}^{(N)}=&\, (a_2^{\dagger})^{c_2}|(z_1,0)\ra_{[c_1]}^{(N-c_2)}\,,\nonumber\\
	\lim\limits_{z_1,z_2\to 0}|\zb\ra_{\mathbbm{c}}^{(N)}=&\,|{\scriptstyle n_0=N-c_1-c_2,\,n_1=c_1,\,n_2=c_2}\ra\,,\label{3CAT_limits}
\end{align}
where $a_i^{\dagger}$ are the bosonic creation operators \eqref{UDgen}, and
\begin{align}
	|(0,z_2)\ra_{[c_2]}\propto&\,\Pi_{[c_2]}|(0,z_2)\ra\nonumber\\
	=&\,2^{-1}\sum_{b_2\in\{0,1\}}(-1)^{c_2b_2}|(0,(-1)^{b_2}z_2)\ra\,,\nonumber\\
	|(z_1,0)\ra_{[c_1]}\propto&\,\Pi_{[c_1]}|(z_1,0)\ra\nonumber\\
	=&\,2^{-1}\sum_{b_1\in\{0,1\}}(-1)^{c_1b_1}|((-1)^{b_1}z_1,0)\ra\,,\label{3CAT_limitsProj}
\end{align}
are reduced-parity projected U(3) CSs,
according to \eqref{projpar} and \eqref{parityCS}. In the expression \eqref{3CAT_limits}, we have also recovered the superscript $|\zb\ra^{(N)}$ notation of the DSCSs \eqref{cohD} to highlight that, the $\mathbbm{c}$-3CAT $|\zb\ra_{\mathbbm{c}}^{(N)}=|(z_1,z_2)\ra_{[c_1,c_2]}^{(N)}$ of $N$ particles, becomes a reduced $[c_2]$-3CAT $|(0,z_2)\ra_{[c_2]}^{(N-c_1)}$ (resp.  $[c_1]$-3CAT $|(z_1,0)\ra_{[c_1]}^{(N-c_2)}$) with $N-c_1$ (resp. $N-c_2$) particles after the limit $z_1\to 0$ (resp. $z_2\to 0$). The new $[c_i]$-3CATs after the limits have a smaller parity symmetry group, as $[c_1]$ and $[c_2]$ belong to $\mathbb{Z}_2^{1}\not=\mathbb{Z}_2^{2}$, the original 3CAT parity group $\mathbb{Z}_2^{D-1}$ for $D=3$. Despite the states in eq.\eqref{3CAT_limitsProj} have a similar structure to the 2CATs in \eqref{eoCS}, they are actually $\mathbb{Z}_2^{1}$-parity adapted U(3)-spin CSs, as they belong to a 3-level Fock space. Furthermore, they have a similar structure to the photon-added CSs, which are defined as a creation operator acting on a canonical CS \cite{PhysRevA.43.492}, but for the U(3)-spin CSs in our case. The photon-added CSs has also been extended to SU(2) \cite{Berrada2015} and SU(1,1) \cite{Monir2019}. As these states have only been studied for the Heisenberg-Weyl group, and for SU(2) \cite{Berrada2015} and SU(1,1) \cite{Monir2019}, the generalization to SU($D$) presents a novel research topic \cite{DcatDecompositionArxiv}.

The $z_i\to 0$ limits in the general $\mathbbm{c}$ and $D$ cases of a $\mathbbm{c}$-DCAT are not straightforward to compute analytically (see \cite{DcatDecompositionArxiv}), thus, the Appendix \ref{App:DCATlimits} is devoted to show in detail these calculations. However, it is necessary to introduce the following limit and notation to progress in our discussion. The zero limit $z_i\to 0$ can be used repeatedly for a set of $l=D-1-k$ different coordinates $\zb_L=\{z_{i_1},\ldots,z_{i_l}\}$, whose indexes are taken from a set of non-repeated indexes $L=\{i_1,\ldots,i_l\}$. Equivalently, we can define the set of the $k$ non-zero coordinates $\zb_K=\{z_{j_1},\ldots,z_{j_k}\}$ which are not used in the limits, where the indexes $K=\{j_1,\ldots,j_k\}$ are not duplicated neither. Note that $\zb=(\zb_K,\zb_L)=(z_1,\ldots,z_{D-1})$ include all the projective coordinates as $k+l=D-1$ by definition. After the limits, the $\mathbbm{c}$-DCAT is transformed into
\begin{equation}
	\lim\limits_{z_L\to \bm{0}_L}\hspace{-1mm}|\zb\ra_{\mathbbm{c}}^{(N)}\hspace{-0.5mm}=\hspace{-0.5mm}(a_{i_1}^{\dagger})^{c_{i_1}}\hspace{-1mm}\ldots\hspace{-0.5mm} (a_{i_l}^{\dagger})^{c_{i_l}}|(\zb_K,\zb_L=\bm{0}_L)\ra_{\mathbbm{c}_K}^{(N-\|\mathbbm{c}_L\|_0)},\label{cDCATs_limits2}
\end{equation}
obtaining a reduced $\mathbbm{c}_K$-DCAT $|(\zb_K,\zb_L=\bm{0}_L)\ra_{\mathbbm{c}_K}$, with $N-\|\mathbbm{c}_L\|_0$ particles, $\mathbbm{c}_K=[c_{j_1},\ldots,c_{j_k}]\in\mathbb{Z}_2^{k}$ parity, and to which it is added a series of $\|\mathbbm{c}_L\|_0$ particles occupying the levels $n_{i_1}=c_{i_1},\ldots,n_{i_l}=c_{i_l}$. The expression $\|\mathbbm{c}_L\|_0$ means the 0-norm (number of non-zero components) of $\mathbbm{c}_L=[c_{i_1},\ldots,c_{i_l}]$. The rest of the notation in \eqref{cDCATs_limits2} is similar to the one used in the eq.\eqref{symDCAT_limits2}. The eq.\eqref{cDCATs_limits2} generalizes the results for $D=2$ in \eqref{2CAT_limits} and for $D=3$ in \eqref{3CAT_limits}.

We will use the equations (\ref{S3C},~\ref{S3CN},~\ref{3CAT_limits}) in Section \eqref{LMGsec} to restore the parity $\mathbbm{c}=\mathbb{0}=[0,0]$ of the variational DSCS 
of a $N$ atoms LMG model with  $D=3$ levels, since the true ground state of this model exhibits a Schr\"odinger cat structure with totally even parity $\mathbb{0}$. We will also see that the other parities in \eqref{3CAT_limits} can model some of the first excited states in the LMG U(3) model. But before that, we shall introduce the Husimi function and some localization measures in phase space to characterize the different quantum phases that appear in the LMG model.

\section{Husimi function and localization measures in phase space}\label{Husimisec}

Coherent states provide  phase space representations (also known as Bargmann/holomorphic representation) of wave functions in quantum physics.  
Here we shall concentrate on the 
Husimi or $Q$-function  \cite{husimi1940} of a pure state $|\psi\ra$, defined as $Q_\psi(z)=|\la z|\psi\ra|^2$ for a given overcomplete set of coherent states $|z\ra$. 
The most popular case 
is in quantum optical systems, for which $|z\ra$ makes reference to a Glauber \cite{PhysRev.131.2766} or canonical (harmonic oscillator) coherent state associated to the Heisenberg-Weyl group. 
This definition can be extended to other coherent state systems 
like those associated to more general symmetry groups  \cite{Perelomov} (see also  \cite{Sugita2002} for some generalizations). 
In our case, the Husimi function of the quantum state \eqref{psisym} is defined in terms of the DSCS coefficients \eqref{coefCS} as 
\be\label{HusimiDef}
Q_\psi(\zb)=|\la\zb|\psi\ra|^2=\left|\sum_{\|\vec{n}\|_1=N}\,\overline{c_{\nb}(\zb)}c_{\vec{n}}(\psi)\right|^2,
\ee
and it is normalized
\be
\int_{\mathbb{C}^{D-1}} Q_\psi(\zb) d\mu(\zb)=1,
\ee
according to the measure \eqref{resounity}. This definition is straightforwardly extended to non pure states defined by a density matrix $\rho$ as $Q_{\rho}(\zb)=\la \zb|\rho|\zb\ra$ (see e.g., \cite{Florian.PhysRevA.69.022317,Sugita2002}).


The Husimi function of a DSCS $|\zb\ra$ is simply $Q_{|\zb\ra}(\zb')=|\la \zb'|\zb\ra|^2$, where  the coherent state overlap $\la \zb'|\zb\ra$ is given in \eqref{scprod}. A more interesting  example is  the Husimi function of a $\mathbbm{c}$-DCAT state $|\zb\ra_{\mathbbm{c}}$ \eqref{DCAT}, which adopts the form 
\begin{align}\label{HusimiDCAT}
	Q_{|\zb\ra_{\mathbbm{c}}}(\zb')=&\,|\la\zb'|\zb\ra_{\mathbbm{c}}|^2=\frac{4^{1-D}}{\mathcal{N}(\zb)_{\mathbbm{c}}^{2}}\left|\sum_{\mathbbm{b}}(-1)^{\mathbbm{c}\cdot\mathbbm{b}}\langle \zb'|\zb^\mathbbm{b}\ra\right|^2\nonumber\\
	=&
	\frac{2^{1-D}\left|\sum_{\mathbbm{b}}(-1)^{\mathbbm{c}\cdot\mathbbm{b}}(1+\zb^\dag\zb'^{\mathbbm{b}})^N\right|^2}
	{(1+\zb'^\dag\zb')^N\sum_{\mathbbm{b}}(-1)^{\mathbbm{c}\cdot\mathbbm{b}}(1+\zb^\dag\zb^{\mathbbm{b}})^N},
\end{align}
where we have used  the coherent state overlap $\langle \zb'|\zb^\mathbbm{b}\ra$ in \eqref{scprod} and the normalization constant $\mathcal{N}(\zb)_{\mathbbm{c}}$ in \eqref{normcat}. 
There are studies in the literature relating the distribution of zeros in phase space of the Husimi function of the ground state of a critical quantum system and the onset of quantum chaos 
(see e.g. \cite{Leboeuf_1990,arranz2013}) and also studies on the  critical behavior of excited states and its relation to order and chaos  \cite{relano}.  Note that, for $\mathbbm{c}$-DCAT  states $|\zb\ra_{\mathbbm{c}}$, the structure of zeros of their Husimi function \eqref{HusimiDCAT} depends on the parity $\mathbbm{c}$. Moreover, the case $D>2$ is much richer and opens new possibilities since $Q$ is multivariate and its zeros are not necessarily isolated points but form curves, surfaces, etc.


In order to visualize the QPT in the critical LMG model across the phase diagram, we shall use the $\nu$-th moments of the Husimi quasi-distribution function 
\be
M_\nu(\psi)=\int_{\mathbb{C}^{D-1}} [Q_\psi(\zb)]^\nu d\mu(\zb), \quad \nu>1.\label{momentsNu}
\ee
Among all Husimi moments, we shall single out $\nu=2$, which corresponds with the so called ``Inverse Participation Ratio'' (IPR) \cite{Aulbach_2004,Calixto_2018}, that measures 
the localization of (inverse area occupied by) $Q_\psi$ in phase space and can be generalized to any probability density function \cite{Wegner1980,Kramer_1993}.
The $\nu$-th moments of the Husimi function supposedly reach their maximum value when $\psi$ itself is a coherent (highly localized) state, that is, when 
$\psi$ only participates of a single coherent state. These conjecture has been proved in the cases of families of coherent states of compact semisimple Lie groups \cite{Sugita2002}, including the symmetric and antisymmetric representations of $SU(D)$ as particular cases \cite{Sugita2003}. This affirmation is widely known as part of the Lieb conjecture, which is mentioned at the end of this subsection. For example, for the particular case of $|\psi\ra=|\zb=\bm{0}\ra=\tfrac{1}{\sqrt{N!}}(a_0^\dag)^N|\vec{0}\ra$ (a boson condensate of $N$ atoms in their ground state $i=0$) and a generic number of levels $D$, a quite straightforward calculations gives 
\begin{align}\label{mumomentD3}
	M_\nu(|\bm{0}\ra)=&\,\frac{(N\nu)!}{N!}
	\frac{(N+D-1)!}{(N\nu+D-1)!}\\ 
	=&\,\frac{(N+D-1)_{D-1}}{(N\nu+D-1)_{D-1}}\xrightarrow{N\to\infty} 1/\nu^{D-1},\nonumber
\end{align}
where $(x)_n=x(x-1)\ldots(x-n+1)$ denotes the descending factorial or Pochhammer symbol. 
The last result \eqref{mumomentD3} can  be straightforwardly extended to any DSCS, that is 
\be\label{mumomentD3CS}
 M_\nu(|\zb\ra)=M_\nu(|\bm{0}\ra), \; \forall \zb\in \mathbb{C}^{D-1},
 \ee
and, in fact, to any boson condensate of $N$ atoms in any level $i=0,\dots,D-1$ (see Appendix \ref{app1} for a proof). 
This in particular means that all coherent states occupy the same area in phase space. 
Indeed, any DSCS $|\zb\ra$ can be obtained by translating/rotating $|\zb=\bm{0}\ra\to U(\zb)|\bm{0}\ra$ by a unitary 
transformation $U(\zb)\in\rmu(D)$ (that is, $|\zb\ra$ can be seen as a ``displaced ground state''), which means that $Q_{|\zb\ra}(\zb')=Q_{|\bm{0}\ra}(\zb'*\zb^{-1})$ with $U(\zb'*\zb^{-1})=U^\dag(\zb)U(\zb')$ the composition of two $\rmu(D)$ transformations; the fact that the Fubini-Study measure $d\mu(\zb)$ in \eqref{resounity} is $\rmu(D)$-invariant completes the proof. Therefore, the $\nu$-moments of the Husimi function of a DSCS $|\zb\ra$ do not depend on the phase space points $\zb\in\mathbb{C}^{D-1}$, but just on $\nu$, the number of particles/atoms $N$, and the number of atom levels $D$. The equations (\ref{mumomentD3},~\ref{mumomentD3CS}) agree with those of Refs. \cite{Giovannetti.PhysRevA.70.022328,PhysRevA.86.032508} in the particular cases of $D=2$ and $D=3$ respectively, and with \cite{Sugita2003} in the general $D$ case. 

The $\mathbbm{c}$-DCAT states  $|\zb\ra_\mathbbm{c}$ in \eqref{DCAT} participate on several coherent states $|\zb^\mathbbm{b}\ra$ and therefore have a lower IPR value (i.e., they occupy a bigger area in phase space), usually a fraction of $M_\nu(|\bm{0}\ra)$. More concretely, the $\nu$-moment of $Q_{|\zb\ra_\mathbbm{c}}$ can be explicitly calculated as in the reference \cite{Sugita2003}, 
\begin{equation}\label{mumomentDCAT}
	M_\nu(|\zb\ra_\mathbbm{c})=M_\nu(|\zb\ra)\sum_{|\vec{k}|=N\nu}|B_{\vec{k}}^2|\,,
\end{equation}
with
\begin{align}
	 B_{\vec{k}}=&\,\sqrt{\frac{(N!)^{\nu}}{(N\nu)!}}\sum_{|\vec{n}_1|=\ldots=|\vec{n}_{\nu}|=N}\binom{\vec{k}}{\vec{n}_1,\vec{n}_2,\ldots,\vec{n}_{\nu}}^{1/2}\nonumber\\
	 &\,\times c_{\nb_1}(\zb)_\mathbbm{c}c_{\nb_2}(\zb)_\mathbbm{c}\cdots c_{\nb_{\nu}}(\zb)_\mathbbm{c}\,,
\end{align}
where $c_{\nb_i}(\zb)_\mathbbm{c}$ are the  $\mathbbm{c}$-DCAT coefficients in the Fock basis \eqref{coefDCAT}. The last sum is restricted to $\nb_1+\nb_2+\ldots+\nb_{\nu}=\vec{k}$, and we are denoting 
\begin{equation}
	\binom{\vec{k}}{\vec{n}_1,\vec{n}_2,\ldots,\vec{n}_{\nu}}\equiv\frac{\vec{k}!}{\vec{n}_1!\cdots\vec{n}_{\nu}!}\,,
\end{equation}
where all the vectors $\nb_1,\nb_2,\ldots,\nb_{\nu}$ correspond to different Fock vectors according to \eqref{symmetricbasis}, i.e. $|\nb_i\ra=|n_{i,0},n_{i,1},\ldots,n_{i,D-1}\ra$, so that we mean by $\vec{n}_i!\equiv \prod_{j=0}^{D-1}(n_{i,j})!\,$ and by $|\vec{n}_i|\equiv \sum_{j=0}^{D-1}n_{i,j}\,$.

In the thermodynamic $N\to\infty$ limit, the bulky expression \eqref{mumomentDCAT} reduces to the more compact one (see Appendix \ref{app2} for a proof)
\begin{align}\label{mumomentDCATlimit}
	\lim\limits_{N\to\infty}M_\nu(|\zb\ra_\mathbbm{c})=(2^{D-1})^{1-\nu}\lim\limits_{N\to\infty}M_\nu(|\zb\ra)=\frac{(2^{D-1})^{1-\nu}}{\nu^{D-1}}\,,
\end{align}
which proves that $\mathbbm{c}$-DCATs  have lower IPR value than DSCSs, since $(2^{D-1})^{1-\nu}<1$ for all $\nu\geq 2$. Hence, DCATs are less localized (occupy a greater area) than DSCSs in phase space. In addition, the limit is independent of the DCAT parity $\mathbbm{c}$. To be more  precise, the equation above is only valid when all the coordinates $z_i$ are non-zero, i.e. $z_i\neq0$ $\forall i=1,\dots,D-1$. Nevertheless, for a totally even $\mathbb{0}$-DCAT which has only $k<D-1$ non-zero vector components  in $\zb$, we can apply the equation \eqref{symDCAT_limits2} for all the $z_i$ that tend to 0, transforming the $\mathbb{0}$-DCAT into a reduced $\mathbb{0}_K$-DCAT with a parity symmetry described by $\mathbb{Z}_2^{k}$. This leads to a expression similar to \eqref{mumomentDCATlimit},  
\begin{align}\label{mumomentDCATlimit2}
	\lim\limits_{N\to\infty}\lim\limits_{\zb_L\to \bm{0}_L}M_\nu(|\zb\ra_{\mathbb{0}})=&\,(2^{k})^{1-\nu}\lim\limits_{N\to\infty}\lim\limits_{\zb_L\to \bm{0}_L}M_\nu(|\zb\ra)\nonumber\\
	=&\,\frac{(2^{k})^{1-\nu}}{\nu^{D-1}}\,,
\end{align}
where the notation is the same as in the eq.\eqref{cDCATs_limits2}.
Note that the denominator $\nu^{D-1}$ is the same as in the equation \eqref{mumomentDCATlimit}, as we calculate the $N\to\infty$ limit of a U($D$) CS, not a U($k$) one (this result is proven in the Appendix \ref{app2}). For a general parity $\mathbbm{c}$-DCAT, the expression above transforms into 
\begin{align}\label{mumomentcDCATlimit2}
	\lim\limits_{N\to\infty}\lim\limits_{\zb_L\to \bm{0}_L}M_\nu(|\zb\ra_{\mathbbm{c}})=\frac{(2^{k+\|\mathbbm{c}_L\|_0})^{1-\nu}}{\nu^{D-1}}\,,
\end{align}
where $\|\mathbbm{c}_L\|_0$ and $k=\|\zb\|_0$ are the number of non-zero components in $\mathbbm{c}_L$ and $\zb$ respectively. The sum $k+\|\mathbbm{c}_L\|_0$ coincides with the number of humps displayed by the Husimi function $Q_{|\zb\ra_{\mathbbm{c}}}(\zb')$ in the phase space coordinates $\zb'$, as we will see in  Section \ref{Sec:Localiz_measures}. The equation \eqref{mumomentcDCATlimit2} includes the eq.\eqref{mumomentDCATlimit2} as a particular case, since $\|\mathbbm{c}_L\|_0=0$ for the $\mathbb{0}$-DCAT.

Instead of $M_\nu(\psi)$, it is sometimes preferred to express delocalization (as a measure of area in phase space) in terms of R\'enyi-Wehrl entropy, which is defined as \cite{Florian.PhysRevA.69.022317,Gnutzmann_2001,Giovannetti.PhysRevA.70.022328} as 
\be\label{RW_entropy}
\mathcal{S}_{W,\nu}(\psi)=\frac{1}{1-\nu}\ln[M_\nu(\psi)],\quad \nu\not=1.
\ee
Taking the limit $\nu\to 1$ in the R\'enyi-Wehrl entropy \eqref{RW_entropy}, one  obtains the Wehrl entropy \cite{Wehrl1978.RevModPhys.50.221} given by
\be\label{WehrlIntegral}
\mathcal{S}_W(\psi)=-\int_{\mathbb{C}^{D-1}} Q_\psi(\zb)\ln[Q_\psi(\zb)] d\mu(\zb).
\ee
Since the  definition of the Husimi function is related to a specific classical phase space (the $\mathbb{C}P^{D-1}$ complex projective space defined by DSCSs in our case), the Wehrl entropy is also called (semi)classical entropy \cite{Wehrl1978.RevModPhys.50.221,lieb2014}. It is the Gibbs entropy continuous form of the Husimi probability function $Q_\rho$ for the state described by a density matrix $\rho$ \cite{Wehrl1978.RevModPhys.50.221,Florian.PhysRevA.69.022317}.
This picture contrast with other common entropies such as the von Neumann entropy $\mathcal{S}_N=-\tr(\rho\ln\rho)$, which we have previously used to study entanglement (quantum nonlocality) in symmetric 
multiquDit systems  \cite{QIP-2021-Entanglement},
and has no immediate relation to classical mechanics. The last one measures how much a state is mixed (non pure), rather than its localization in phase-space. According to the Lieb conjecture \cite{Lieb1978}, the minimum Wehrl entropy \eqref{WehrlIntegral} is attained when $\psi$ is a DSCS. It was proved for SU(2) spin-$j$ CSs in \cite{lieb2014,ANNAthesis}, for symmetric SU($D$) spin CSs (DSCSs for us) in \cite{Lieb2016}, and for any compact semisimple Lie group in \cite{Sugita2002}. The minimum Wehrl entropy value can be easily obtained from the Husimi $\nu$-moment of the highest-weight vector $|\bm{z}\ra=|\bm{0}\ra$ in \eqref{mumomentD3}, once we realize that 
$\mathcal{S}_W=\lim_{\nu\to 1}\mathcal{S}_{W,\nu}$ in \eqref{RW_entropy}, and that 
\be
\min_{\psi}\mathcal{S}_W(\psi)=\mathcal{S}_W(|\zb\ra)=\mathcal{S}_W(|\bm{0}\ra)
\ee
according to \eqref{mumomentD3CS}. Therefore, taking the limit $\nu\to 1$ in \eqref{mumomentD3} we arrive to 
\begin{align}
	\label{Wehrl_entropy_CS_Nlimit}
	\mathcal{S}_W(|\bm{0}\ra)=&\,N\big(\psi^{(0)}(N+D)-\psi^{(0)}(N+1)\big)\\
	=&\, N\sum_{k=1}^{D-1}\frac{1}{N+k}\xrightarrow{N\to\infty} D-1, \nonumber	
\end{align}
for a generic number $D$ of levels, where $\psi^{(0)}(x)=\Gamma'(x)/\Gamma(x)$ is the digamma function.
There is a particular version of this result for a $U(3)$ vibron model in \cite{PhysRevA.86.032508}. As a particular case, in the thermodynamic limit $N\to\infty$, the minimum Wehrl entropy  
is $D-1=1$ for $D=2$, which is the minimum value of the Wehrl entropy predicted by Lieb in \cite{Lieb1978} for the harmonic oscillator coherent states (Heisenberg-Weyl group). This is so because  Bloch SU(2) spin-$j$ coherent states $|z\ra$  tend to the Heisenberg-Weyl (harmonic oscillator)  coherent states $|\alpha\ra$ in the large spin limit $j=N/2 \gg 1$ with the rescaling $z=\alpha/\sqrt{N}$ \cite{Perelomov,Lieb1973}. Unlike for DSCSs, we do not have closed analytical formulas for the Wehrl entropy of $\mathbbm{c}$-DCAT states, except in the thermodynamic limit when, in general, 
\begin{align}\label{WehrlEntDCATlimit}
	\lim\limits_{N\to\infty}\mathcal{S}_W(|\zb\ra_{\mathbbm{c}})=&\,\lim\limits_{N\to\infty}\mathcal{S}_W(|\zb\ra)+\log(2^{D-1})\nonumber\\
	=&\,(D-1)(1+\log(2)),
\end{align}
being the same for all different parities $\mathbbm{c}$. As we already commented in the equation \eqref{mumomentDCATlimit2}, when there are only $k$ non-zero components in $\zb$ for the fully even $\mathbb{0}$-DCAT, the expression above \eqref{WehrlEntDCATlimit} has to be replaced by
\begin{equation}\label{WehrlEntDCATlimit2}
	\lim\limits_{N\to\infty}\lim\limits_{\zb_L\to\bm{0}_L}\mathcal{S}_W(|\zb\ra_{\mathbb{0}})
	=(D-1)+k\log(2)\,.
\end{equation}
Therefore, the totally even parity adaptation of a DSCS entails a Wehrl entropy (area in phase space) excess of
\begin{equation}\label{WehrlEntDCATlimitDSCS}
\mathcal{S}_W(|\zb\ra_{\mathbb{0}})-\mathcal{S}_W(|\zb\ra)\xrightarrow{N\to\infty} k\log(2)
\end{equation}
in the thermodynamic limit.
 This is a particular case of the result proposed by Mintert and Zyczkowski in \cite{Florian.PhysRevA.69.022317}. Also, the limits (\ref{mumomentDCATlimit2},\ref{WehrlEntDCATlimit2}) for the $\mathbb{0}$-DCAT  generalize the results obtained in \cite{PhysRevA.86.032508} for $D=3$ and $\zb=(z_1,-\bar{z}_1)$, which is equivalent to have only one non-zero component in $\zb$, i.e., $k=1$. For the general $\mathbbm{c}$-parity case, we use the eq.\eqref{mumomentcDCATlimit2} to obtain 
 \begin{equation}\label{WehrlEntcDCATlimit2}
 	\lim\limits_{N\to\infty}\lim\limits_{\zb_L\to\bm{0}_K}\mathcal{S}_W(|\zb\ra_{\mathbbm{c}})
 	=(D-1)+(k+\|\mathbbm{c}_L\|_0)\log(2)\,.
 \end{equation}
All the expressions in the thermodynamic limit presented in this section are examined in more detail in the Appendix \ref{app2}.


In the section \ref{Sec:Localiz_measures}, we propose Husimi second moments and R\'enyi-Wehrl entropies of the ground state of a $3$-level atom LMG model \eqref{hamUDbis} as localization measures in phase space, in order to characterize the three quantum phases appearing in this model. But previously we are going to study the phase diagram of the critical $D=3$ level  LMG model   in the next section.

\section{LMG model for three-level atoms and its quantum phase diagram} \label{LMGsec}

%

We particularize the Hamiltonian \eqref{hamUDbis} for $D=3$ (3-level atoms or qutrits). Therefore, our Hamiltonian density will be
\begin{equation}
H=\frac{\epsilon}{N}(S_{33}-S_{11})-\frac{\lambda}{N(N-1)}\sum_{i\not=j=1}^3 S_{ij}^2.\label{hamU3}
\end{equation}
We shall measure  energy in $\epsilon$ units  and discuss the energy spectrum and the phase diagram in terms of the only control parameter $\lambda$. 
In \cite{nuestroPRE} we have proved that this model displays three different quantum phases for the completely symmetric unitary irreducible representation of U(3) labelled by the total number of particles $N$; Ref. \cite{nuestroPRE} also studies other permutation symmetry sectors (fermionic mixtures from two-row Young diagrams) which will not be discussed here. Let us summarize the essential points. Coherent (semiclassical) states are in general good 
variational states which faithfully reproduce the ground state energy of Hamiltonian models in the semiclassical/thermodynamic  limit $N\to\infty$. Therefore, we define 
the energy surface associated to the Hamiltonian density $H$ in \eqref{hamU3}  as the DSCS expectation value of the Hamiltonian density in the thermodynamic limit 
\begin{align}\label{energyU3}
 E_{|\zb\ra}(\epsilon,\lambda)=&\lim_{N\to\infty}\langle \zb|H|\zb\rangle\\
 =&\lim_{N\to\infty}\hspace{-1mm}\left(\hspace{-0.5mm}
 \epsilon\tfrac{\la \zb|S_{33}|\zb\ra-\la \zb|S_{11}|\zb\ra}{N}-\lambda\tfrac{\sum_{i\not=j=1}^3 \la \zb|S_{ij}|\zb\ra^2}{N(N-1)}\hspace{-0.5mm}\right)\hspace{-1mm},\nonumber
\end{align}
with $\la \zb|S_{ij}|\zb\ra$ in  \eqref{CSEV}. Note that we have used that there are no spin fluctuations in the thermodynamic limit \eqref{nofluct}.  
Denoting $\zb=(z_1,z_2)$ the phase space coordinates for U(3)-spin coherent states \eqref{cohD}, the energy surface has the explicit form
\be\label{enersym}
	E_{|\zb\ra}(\epsilon,\lambda)= \epsilon\frac{ |z_2|^2-1}{
		|z_1|+|z_2|^2+1}
	-\lambda\frac{ z_1 ^2 \left(\bar{z}_2^2+1\right)+z_2^2+\mathrm{c.c.}
	}{\left(|z_1|+|z_2|^2+1\right)^2}.
\ee
This energy surface is invariant under parity transformations $z_1\to-z_1$, $z_2\to-z_2$, a symmetry  which is inherited from the discrete parity symmetry of the Hamiltonian
\eqref{hamU3}. In fact, the energy surface $E_{|\zb\ra}(\epsilon,\lambda)$ coincides with all $\mathbbm{c}$-DCAT Hamiltonian expectation values in the thermodynamic limit, that is
\be\label{DCATenergysurface} E_{|\zb\ra_\mathbbm{c}}(\epsilon,\lambda)=E_{|\zb\ra}(\epsilon,\lambda) \quad\forall\mathbbm{c}\in\mathbb{Z}_2^{D-1}.
\ee
This can be seen by using the linear and quadratic $\rmu(D)$-spin operator expectation values in a $\mathbbm{c}$-parity DCAT defined in \cite{QIP-2021-Entanglement}, and realizing that 
\be \lim_{N\to\infty} {}_\mathbbm{c}\la\zb|S_{ij}|\zb\ra_\mathbbm{c}=\lim_{N\to\infty}\la\zb|S_{ij}|\zb\ra,
\ee
which can also be extended to quadratic (two-body) $\rmu(D)$-spin operator expectation values because of the absence of quantum fluctuations in the thermodynamic limit \eqref{nofluct}. 
This fact has important consequences in the spontaneous breakdown of the parity symmetry in the thermodynamic limit and the quantum phase transition, as we are going to see in the following. 

The variational minimum (ground state) energy 
\be E^{(0)}(\epsilon,\lambda)=\mathrm{min}_{z_1,z_2\in \mathbb{C}}E_{|(z_1,z_2)\ra}(\epsilon,\lambda)\ee
is attained at the critical (real) coherent state parameters
\bea
z_{1\pm}^{(0)}(\epsilon,\lambda)&=&\pm\left\{\begin{array}{lll}
 0, && 0\leq \lambda \leq \frac{\epsilon }{2}, \\
 \sqrt{\frac{2\lambda- \epsilon }{2 \lambda +\epsilon }}, && \frac{\epsilon }{2}\leq \lambda \leq \frac{3 \epsilon }{2}, \\
 \sqrt{\frac{2\lambda }{2 \lambda +3 \epsilon }}, && \lambda \geq \frac{3 \epsilon }{2},
\end{array}\right.\nonumber\\
z_{2\pm}^{(0)}(\epsilon,\lambda)&=&\pm\left\{\begin{array}{lll}
 0, & & 0\leq \lambda \leq  \frac{3 \epsilon }{2}, \\
 \sqrt{\frac{2 \lambda -3 \epsilon}{2 \lambda +3 \epsilon }}, & & \lambda \geq \frac{3 \epsilon }{2}. \end{array}\right. \label{critalphabeta}
\eea
Inserting \eqref{critalphabeta} into  \eqref{enersym} gives the ground state energy density in the thermodynamic limit 
\be
E^{(0)}(\epsilon,\lambda)=\left\{\begin{array}{lllr}
 -\epsilon,  && 0\leq \lambda \leq \frac{\epsilon }{2}, & \mathrm{(I)}\\
 -\frac{(2 \lambda +\epsilon )^2}{8 \lambda }, && \frac{\epsilon }{2}\leq \lambda \leq \frac{3 \epsilon }{2}, &  \mathrm{(II)} \\
  -\frac{4\lambda^2+3\epsilon ^2}{6 \lambda }, & &\lambda \geq \frac{3 \epsilon }{2}. &  \mathrm{(III)}\end{array}\right.\label{energysym}
\ee
Here we clearly distinguish three different phases: I, II and III, and two second-order QPTs (according to Ehrenfest's classification) occurring at critical points  $\lambda^{(0)}_{\mathrm{I}\leftrightarrow\mathrm{II}}=\epsilon/2$ and 
$\lambda^{(0)}_{\mathrm{II}\leftrightarrow\mathrm{III}}=3\epsilon/2$, respectively, at which the second derivative of $E_0(\epsilon,\lambda)$ is discontinuous. As we have already anticipated, the ground state is degenerated, since there are four different DSCSs $|z_{1\pm}^{(0)},z_{2\pm}^{(0)}\ra$ (or equivalently, four 3CAT states $|\zb\ra_\mathbbm{c}$ with parities $\mathbbm{c}=[0,0],[1,0],[0,1],$ and $[1,1]$) with the same energy \eqref{energysym} in the thermodynamic limit $N\to\infty$. This is a consequence  of the spontaneous breakdown of the discrete parity symmetry $\mathbb{Z}_2^{2}$ of the Hamiltonian \eqref{hamU3}, as was already pointed out in \cite{nuestroPRE}. For general $D$, the ground state degeneracy would go as $2^{k}$, with $k$ the number of nonzero components of $\zb^{(0)}$, with a maximum degeneracy of $2^{D-1}$ (the number of elements of the parity group $\mathbb{Z}_2^{D-1}$).

\section{Fidelity between variational cats and numerical low-lying Hamiltonian eigenstates with definite parity}\label{Sec:Fidelity}

\begin{figure}[h]
	\centering
	\includegraphics[width=0.5\textwidth]{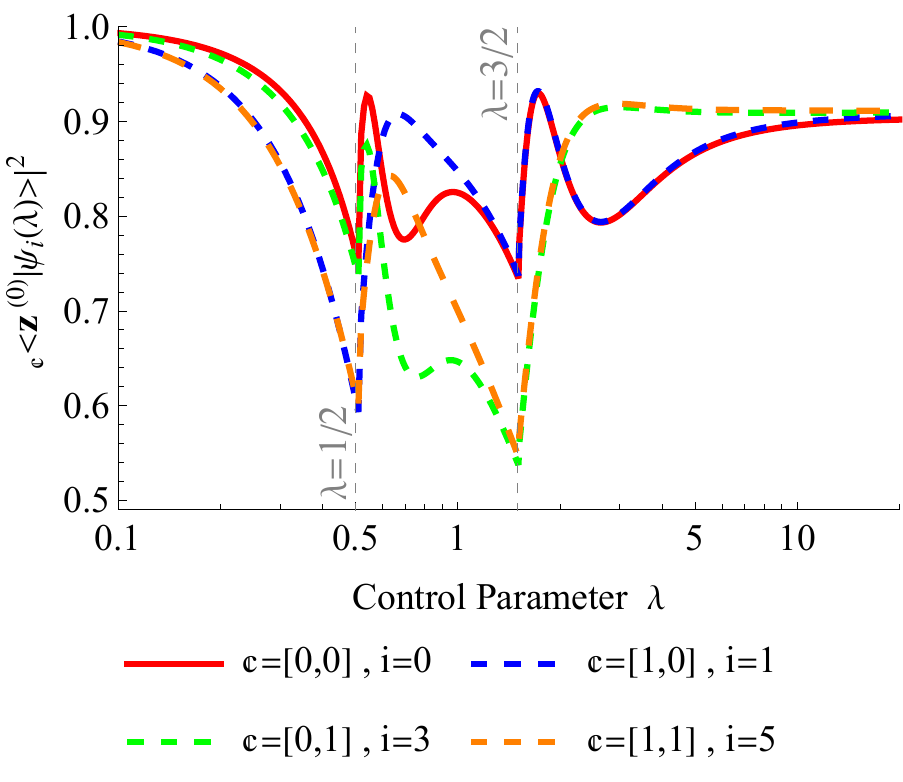}
	\caption{Fidelity $|_\mathbbm{c}\la \zb^{(0)}|\psi_i(\lambda)\ra|^2$ between the variational $\mathbb{c}$-3CATs \eqref{varES} and the  numerical LMG Hamiltonian eigenstates $\psi_i$ as a function of $\lambda$ ($\epsilon$ units and log-scale in abscissa axis). Vertical grid lines denote the quantum critical points.}
	\label{Figure:Fidelityfig1}
\end{figure}

For a finite number $N$ of atoms, coherent states $|\zb\ra$ still provide a fairly good approximation to the ground state when properly adapted to the (not yet broken) parity. There are two possible variational approaches for finite $N$: 
\begin{enumerate}
 \item Project $|\zb\ra$ onto parity $\mathbbm{c}=\mathbb{0}=[0,0]$ (the ground state is always totally even), use this $\mathbb{0}$-3CAT state $|\zb\ra_\mathbb{0}$ as a variational state, and determine the critical 
coherent state parameters $\zb^{(0,N)}$ that minimize the energy expectation value $_\mathbb{0}\la \zb|H|\zb\ra_\mathbb{0}$ for finite $N$ (the matrix elements $_\mathbb{0}\la \zb|S_{ij}|\zb\ra_\mathbb{0}$ can be found in \cite{QIP-2021-Entanglement}), or
\item Use one of the four critical coherent state parameters combinations $\zb^{(0)}=(z_{1+}^{(0)},z_{2+}^{(0)})$ obtained for $N\to\infty$ in \eqref{critalphabeta}, substitute them into $|\zb\ra$ for finite $N$ creating $|\zb^{(0)}\ra$, then restore parity by projecting onto fully even parity 
\begin{align}
	\Pi_\mathbb{0}|\zb^{(0)}\ra=\frac{1}{4}&\left[|z_{1+}^{(0)},z_{2+}^{(0)}\ra+|z_{1+}^{(0)},z_{2-}^{(0)}\ra\right.\nonumber\\
	&\,+\left.|z_{1-}^{(0)},z_{2+}^{(0)}\ra+|z_{1-}^{(0)},z_{2-}^{(0)}\ra\right]\label{varground}
\end{align}
and normalize
\be 
|\zb^{(0)}\ra_{\mathbb{0}}=\frac{\Pi_\mathbb{0}|\zb^{(0)}\ra}{\mathcal{N}(\zb^{(0)})_{\mathbb{0}}}\label{vargroundNorm}.
\ee
\end{enumerate}
The second procedure is less accurate but much easier. We shall adopt it in the following to obtain variational approximations $|\zb^{(0)}\ra_\mathbb{0}$ (the properly normalized 
projection of $\Pi_\mathbb{0}|\zb^{(0)}\ra$) to the ground state $|\psi_0\ra$, and to evaluate how faithful (in the sense of \cite{doi:10.1080/09500349414552171}) they are to numerical solutions obtained by direct Hamiltonian diagonalization. Moreover, we shall naively extend this procedure to evaluate the fidelity between other $\mathbbm{c}$-3CATs $|\zb^{(0)}\ra_\mathbbm{c}\propto\Pi_\mathbbm{c}|\zb^{(0)}\ra $ and the first excited states $|\psi_i\ra, i=1,2,3,4,5$ (in increasing order of energy), which have definite parity $\mathbbm{c}$ and are obtained by numerical diagonalization of the Hamiltonian \eqref{hamU3} for different values of the control parameter $\lambda$. In this case, the $\mathbbm{c}$-3CATs are reduced to a smaller parity group 3CATs when some of the coordinates in $\zb^{(0)}=(z_{1+}^{(0)},z_{2+}^{(0)})$ tend to 0 (see equation \eqref{3CAT_limits} and the discussion below it). Therefore, it would be more precise to define the variational excited states (ES for short) as
\begin{equation}
	|\zb^{(0)}\ra_{\mathbbm{c}}=\lim_{\zb\to\zb^{(0)}}|\zb\ra_{\mathbbm{c}}\,,\quad \forall \mathbbm{c}\neq\mathbb{0}\label{varES}
\end{equation}
rather than directly using  the equation \eqref{vargroundNorm}, in order to avoid a null projection (see the discussion above the eq.\eqref{2CAT_limits} for more details). This will become important when plotting the Figures \ref{Figure:Fidelityfig1}, \ref{Figure:HusimiFunc_c-3CATs_Variational_LMG} and \ref{Figure:IPRc3CATs_NumNVar_LMG}.

\begin{figure}[h]
	\centering
	
	\includegraphics[width=0.5\textwidth]{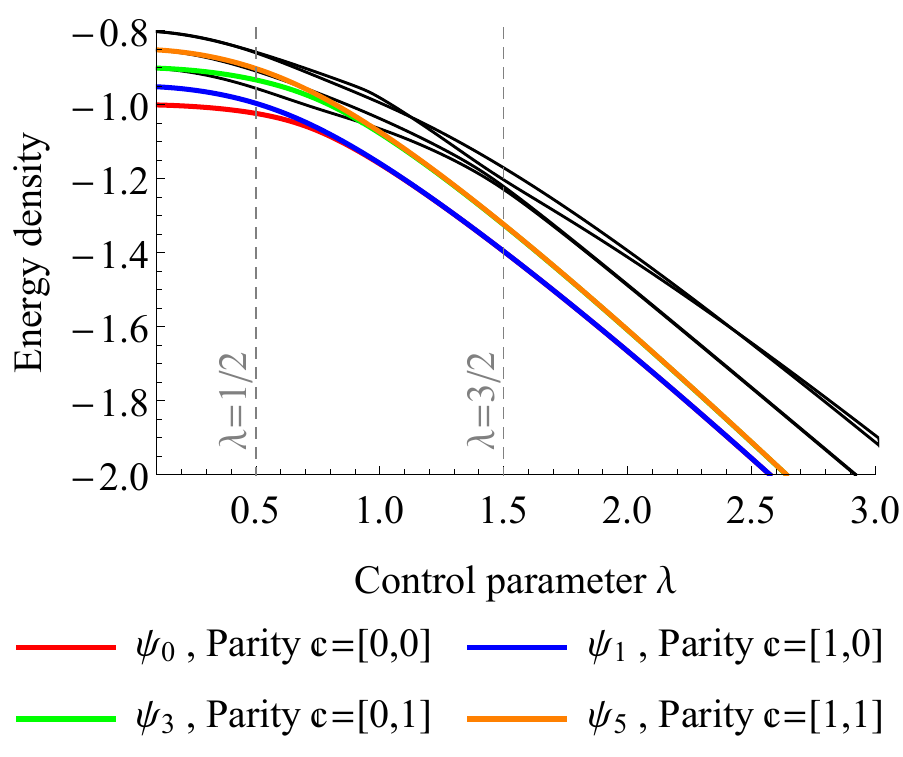}
	\caption{Energy density spectrum  of the first excited states of the LMG $\rmu(3)$ model, obtained by numerical diagonalization of the LMG U(3) Hamiltonian \eqref{hamU3} for $N=20$ particles, as a function of the control parameter $\lambda$. The colored lines represent states with well defined parity, which is indicated in the legend. The quantum critical points $\lambda^{(0)}_{\mathrm{I}\leftrightarrow\mathrm{II}}= \epsilon/2$ and $\lambda^{(0)}_{\mathrm{II}\leftrightarrow\mathrm{III}}= 3\epsilon/2$ are indicated by vertical dashed grid lines. Energies and $\lambda$ are given in $\epsilon$ units.}
	\label{Figure:Energies}
\end{figure}

The condition for a Hamiltonian eigenstate $|\psi_i\ra$ to have a definite parity $\mathbbm{c}$ is $\la\psi_i|\Pi_{\mathbbm{c}}|\psi_i\ra=1$. In particular, for $N=20$  and $\lambda\in(0,3)$, we have obtained the following parities for the fist low-lying Hamiltonian eigenstates (in increasing order of energy) 
\begin{align}\label{OverlapEq1}
	\la\psi_0|\Pi_{[0,0]}|\psi_0\ra=&\,1,\quad
	\la\psi_1|\Pi_{[1,0]}|\psi_1\ra=1,\quad
	\\
	\la\psi_2|\Pi_{[0,0]}|\psi_2\ra=&\,1,\quad
	\la\psi_3|\Pi_{[0,1]}|\psi_3\ra=1,\quad
	\nonumber\\
	\la\psi_4|\Pi_{[1,0]}|\psi_4\ra=&\,1,\quad
	\la\psi_5|\Pi_{[1,1]}|\psi_5\ra=1.\quad
	\nonumber
\end{align}
In Figure \ref{Figure:Energies} we represent the low-lying spectrum of the LMG Hamiltonian \eqref{hamU3} as a function of the control parameter $\lambda$ for $N=20$ particles. 
 The four colored lines represent the states $\psi_i,$ $i=0,1,3,5$ which have the same $\mathbbm{c}$-parity of specific $\mathbb{c}$-DCATs. After the first phase transition around $\lambda^{(0)}_{\mathrm{I}\leftrightarrow\mathrm{II}}= \epsilon/2$, the states $i=0,1$ (red and blue) start getting closer until they finally merge for large $\lambda$. This degeneracy in the ground state for finite $N$ can be considered as a "precursor" of the first QPT at $\lambda=\epsilon/2$. The degeneracy is also present in the excited states $i=3,5$ (green and orange) around $\lambda^{(0)}_{\mathrm{I}\leftrightarrow\mathrm{II}}$. Furthermore, as we move towards the next critical point $\lambda^{(0)}_{\mathrm{II}\leftrightarrow\mathrm{III}}= 3\epsilon/2$, the states $i=0,1,3,5$ start to merge in a 4-fold degenerate ground state, providing another ``precursor" but for the second QPT at $\lambda=3\epsilon/2$. This degeneracy phenomenom is more and more evident as we approach the thermodynamic limit, where the ground state is completely 4-fold degenerate.

Figure \ref{Figure:Fidelityfig1} shows the fidelity 
\be\label{fidEq}
F(|\zb^{(0)}\ra_\mathbbm{c},|\psi_i\ra)=|_\mathbbm{c}\la \zb^{(0)}|\psi_i\ra|^2,
\ee
between variational excited states \eqref{varES} and numerical low-lying Hamiltonian eigenstates $\psi_i$ with the same parity  $\mathbbm{c}$  (states with different parities are orthogonal). As expected, the 3CAT state $| \zb^{(0)}\ra_\mathbb{0}$ gives a fairly good approximation to the ground state $|\psi_0\ra$, with a high fidelity $F\gtrsim 0.8$ (specially in phase I), except near the critical points $\lambda^{(0)}_{\mathrm{I}\leftrightarrow\mathrm{II}}=\epsilon/2$ and $\lambda^{(0)}_{\mathrm{II}\leftrightarrow\mathrm{III}}=3\epsilon/2$, where fidelity always drops. 
Figure \ref{Figure:Fidelityfig1} also shows the fidelity between the variational approximations $| \zb^{(0)}\ra_\mathbbm{c}$, with parities $\mathbbm{c}=[1,0], [0,1], [1,1]$, and the excited states $|\psi_i\ra, i=1, 3, 5$, respectively. The excited states $|\psi_2\ra$ and $|\psi_4\ra$ are not considered in this discussion because they already share parity with $|\psi_0\ra$ and $|\psi_1\ra$, respectively, and therefore they can not be faithfull to $| \zb^{(0)}\ra_{[0,0]}$ and $| \zb^{(0)}\ra_{[1,0]}$ since $\la \psi_2|\psi_0\ra=0$ and $\la \psi_4|\psi_1\ra=0$, i.e., they are  mutually orthogonal as Hamiltonian eigenstates with diferent eigenvalues. 
Let us continue discussing the Figure \ref{Figure:Fidelityfig1}. The fidelity $|_{[1,0]}\la \zb^{(0)}|\psi_1\ra|$ is also fairly high, although not as much as for the ground state. Note that, according to Eq. \eqref{critalphabeta},  the first component $\zb^{(0)}_1$ of $\zb^{(0)}$ is zero in phase I and $\zb^{(0)}_2=0$ in phases I and II. Therefore, according to the equations \eqref{3CAT_limits} and \eqref{3CAT_limitsProj}, in the phases I and II, the fidelity must be calculated using reduced-parity 3CATs. For instance, in the phase I, the 3CAT $|\zb^{(0)}\ra_{[1,0]}$ becomes  a Fock basis state $|{\scriptstyle n_0=N-1,\,n_1=1,\,n_2=0}\ra$ because $\zb^{(0)}(\lambda)=(0,0)$ at $\lambda<\epsilon/2$; and in the phase II, it ``transmutes'' to a $\mathbb{Z}_2$-parity 3CAT $|(z_{1+}^{(0)},0)\ra_{[1]}^{(N)}$.
The same happens with the fidelities $|_{[0,1]}\la \zb^{(0)}|\psi_3\ra|$ and $|_{[1,1]}\la \zb^{(0)}|\psi_5\ra|$, which are fairly high far from the critical points. All the fidelities presented in Figure \ref{Figure:Fidelityfig1} tend to 1 when $\lambda\to 0$, which corresponds to the coordinates $\zb^{(0)}(\lambda)=(0,0)$. This is possible because the numerical diagonalization in the noninteracting case ($\lambda=0$) reproduces very accurately the Fock basis states at the bottom of the eq. \eqref{3CAT_limits}. The spectrum classification of the non-interacting LMG U(3) model was already studied analytically in \cite{nuestroPRE}, giving Fock basis states as eigenstates of the Hamiltonian. Additionally, the 4-fold degeneracy of the eigenstates $i=0,1,3,5$ is present in Figure \ref{Figure:Fidelityfig1} at high  $\lambda\gg 1$, where all the colored lines merge.



\begin{figure}[h]
	\centering
	\includegraphics[width=0.5\textwidth]{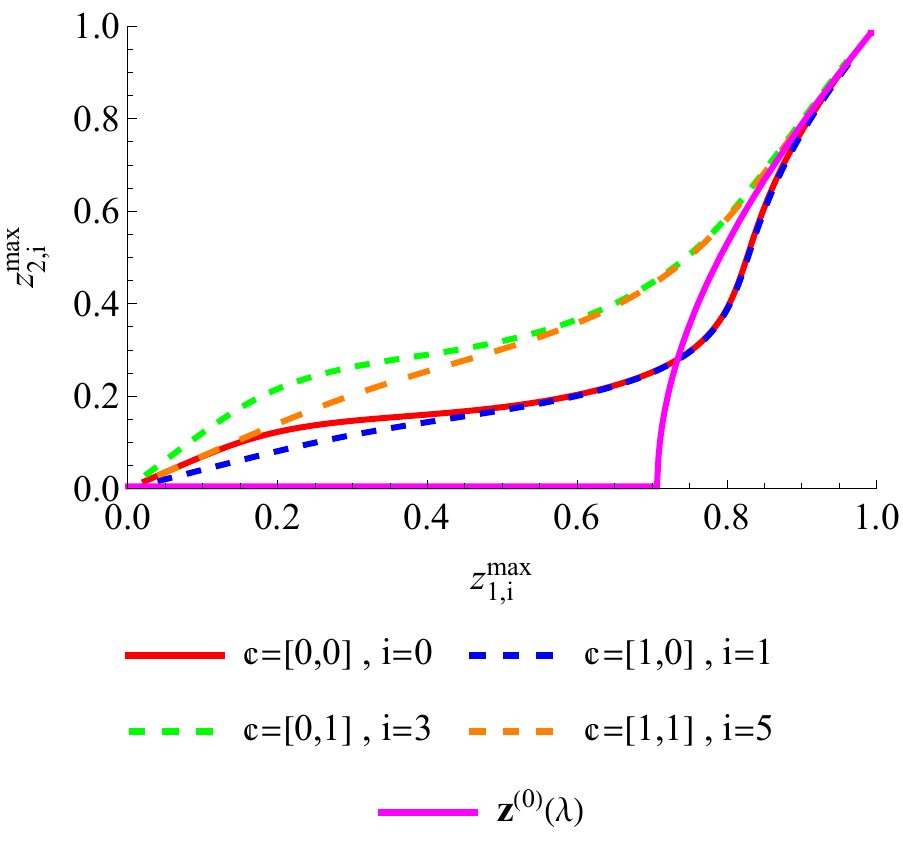}
	\caption{Parametric plot of the fitting points $\zb^{\mathrm{max}}_{i}=(z_{1,i}^{\mathrm{max}},z_{2,i}^{\mathrm{max}})$ maximizing the overlap or fidelity $\left|_{\mathbbm{c}}\la \zb|\psi_i(\lambda)\ra\right|^2$, as a  function of $\lambda\in(0,20)$ ($\epsilon$ units and log-scale) for $N=20$ particles. The fitting points are compared to the critical values $\zb^{(0)}=(z_{1\pm}^{(0)},z_{2\pm}^{(0)})$ in \eqref{critalphabeta}, represented by the solid magenta line.}
	\label{Figure:MaximizationOverlapZcrit}
\end{figure}

\begin{figure}[h]
	\centering	\includegraphics[width=0.5\textwidth]{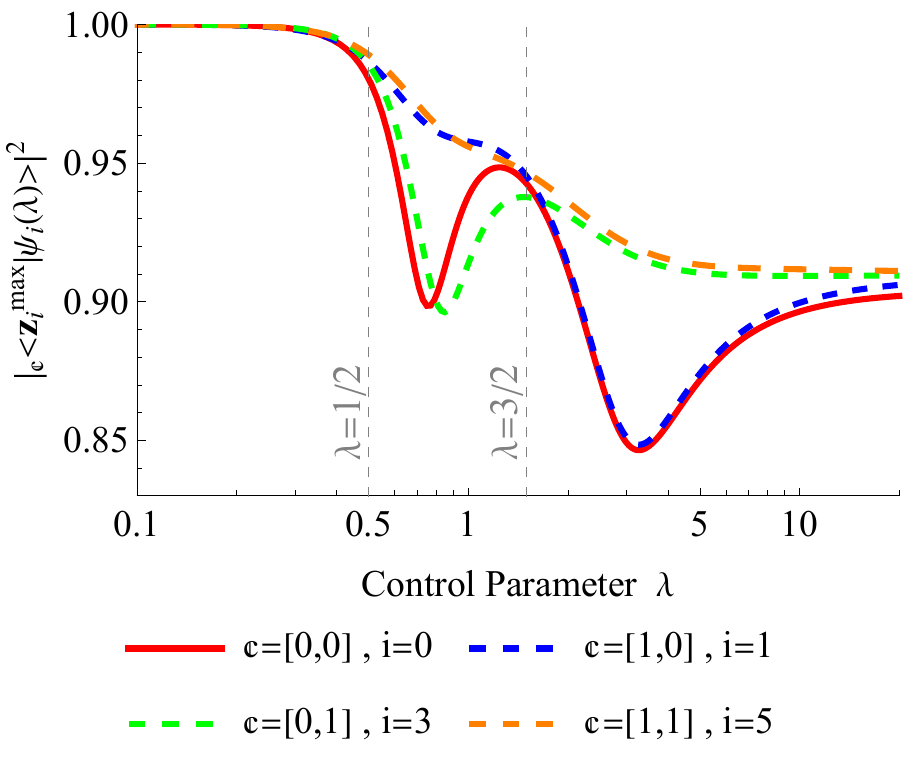}
	\caption{Maximum overlap or fidelity $\left|_{\mathbbm{c}}\la \zb^{\mathrm{max}}_{i}|\psi_i(\lambda)\ra\right|^2$ between the $\mathbbm{c}$-DCATs $|\zb\ra_{\mathbbm{c}}$ and the LMG numerical eigenvectors $|\psi_i(\lambda)\ra$ of different parity as a function of $\lambda$ ($\epsilon$ units and log-scale) for $N=20$ particles.}
	\label{Figure:MaximizationOverlap}
\end{figure}

The failure of the variational state $| \zb^{(0)}\ra_\mathbb{c}$ to properly represent  the numerical Hamiltonian eigenstate   $|\psi_i\ra$  (for the corresponding parity $\mathbb{c}$) near the quantum critical points $\lambda=\epsilon/2$ and $\lambda=3\epsilon/2$, can be fixed by simply  maximizing the overlap 
\be
\left|_{\mathbbm{c}}\la \zb|\psi_i(\lambda)\ra\right|^2=\mathcal{N}(\zb)_\mathbbm{c}^2Q_{\psi_i(\lambda)}(\zb)
\ee
in the phase space coordinates $\zb=(z_1,z_2)$ for each value of $\lambda$. This procedure, of course, results in fitting values $\zb^{\mathrm{max}}_{i}=(z_{1,i}^{\mathrm{max}},z_{2,i}^{\mathrm{max}})$, which are different from the critical values $\zb^{(0)}=(z_{1\pm}^{(0)},z_{2\pm}^{(0)})$ in \eqref{critalphabeta} at the thermodynamic limit. Indeed, in Figure  \ref{Figure:MaximizationOverlapZcrit} we plot the (real) values of $\zb^{\mathrm{max}}_{i}$, to be compared to $\zb^{(0)}$, as a function of $\lambda$. Both values meet at $\lambda=0$ and $\lambda\gg 1$, i.e. when the two-body interaction is not present and when it predominates, respectively. Then, in Figure \ref{Figure:MaximizationOverlap}, we represent the overlap $\left|_{\mathbbm{c}}\la \zb^{\mathrm{max}}_{i}|\psi_i(\lambda)\ra\right|^2$, which now attains values above 0.8 for all values of $\lambda$, thus improving the results of \eqref{fidEq}.

\section{Localization measures of the ground state in phase space throughout the phase diagram}\label{Sec:Localiz_measures}

Now we are interested in analyzing the QPT of the three-level atom LMG model by using the localization measures (area in phase space) introduced in Section \ref{Husimisec}. 

Let us start by analyzing the structure of the Husimi function $Q_{|\zb^{(0)}\ra_{\mathbb{0}}}(\zb')$ of the variational ground state $|\zb^{(0)}\ra_{\mathbb{0}}$ (see eq.\eqref{vargroundNorm}). The variational Husimi function $Q_{|\zb^{(0)}\ra_{\mathbb{0}}}(\zb')$ depends on the complex phase space coordinates  $\zb'=(z_1',z_2')\in \mathbb{C}^2$. It also depends on the control parameter $\lambda$ through the critical point $\zb^{(0)}=(z_{1+}^{(0)},z_{2+}^{(0)})$ (we take $\epsilon$ energy units for simplicity, see eq.\eqref{critalphabeta}). 
In order to plot $Q_{|\zb^{(0)}\ra_{\mathbb{0}}}(\zb')$ in phases I, II and III, we shall separate ``position'' $x_{1,2}=\text{Re}(z'_{1,2})$ and ``momentum''  $p_{1,2}=\text{Im}(z'_{1,2})$ coordinates (see e.g. \cite{Hall1994,Hall1997} for phase-space approaches to quantum mechanics and \cite{PhysRevA.86.032508,real2013} for a justification in other models, like quadratures of the electromagnetic field). 

\begin{figure}[h]
	\begin{center}
		\includegraphics[width=4.2cm]{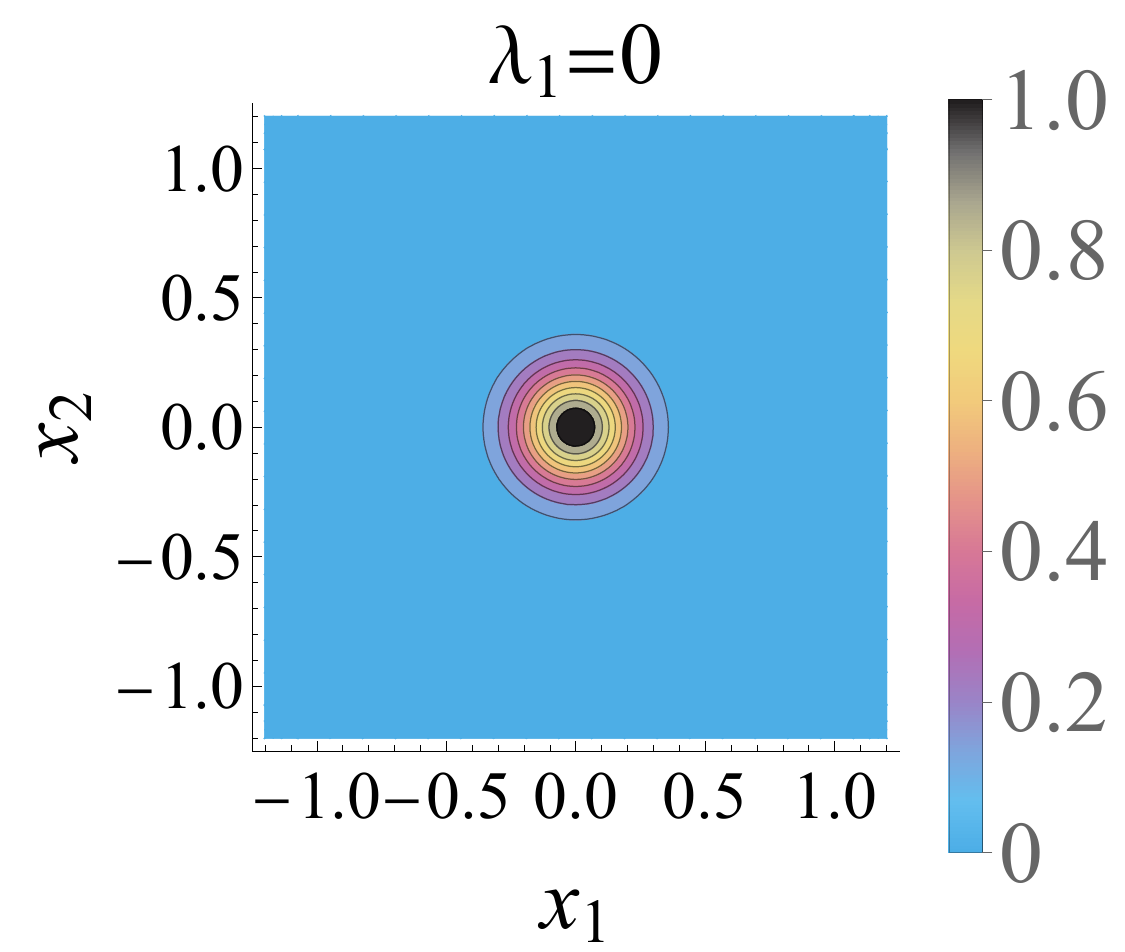}
		\includegraphics[width=4.2cm]{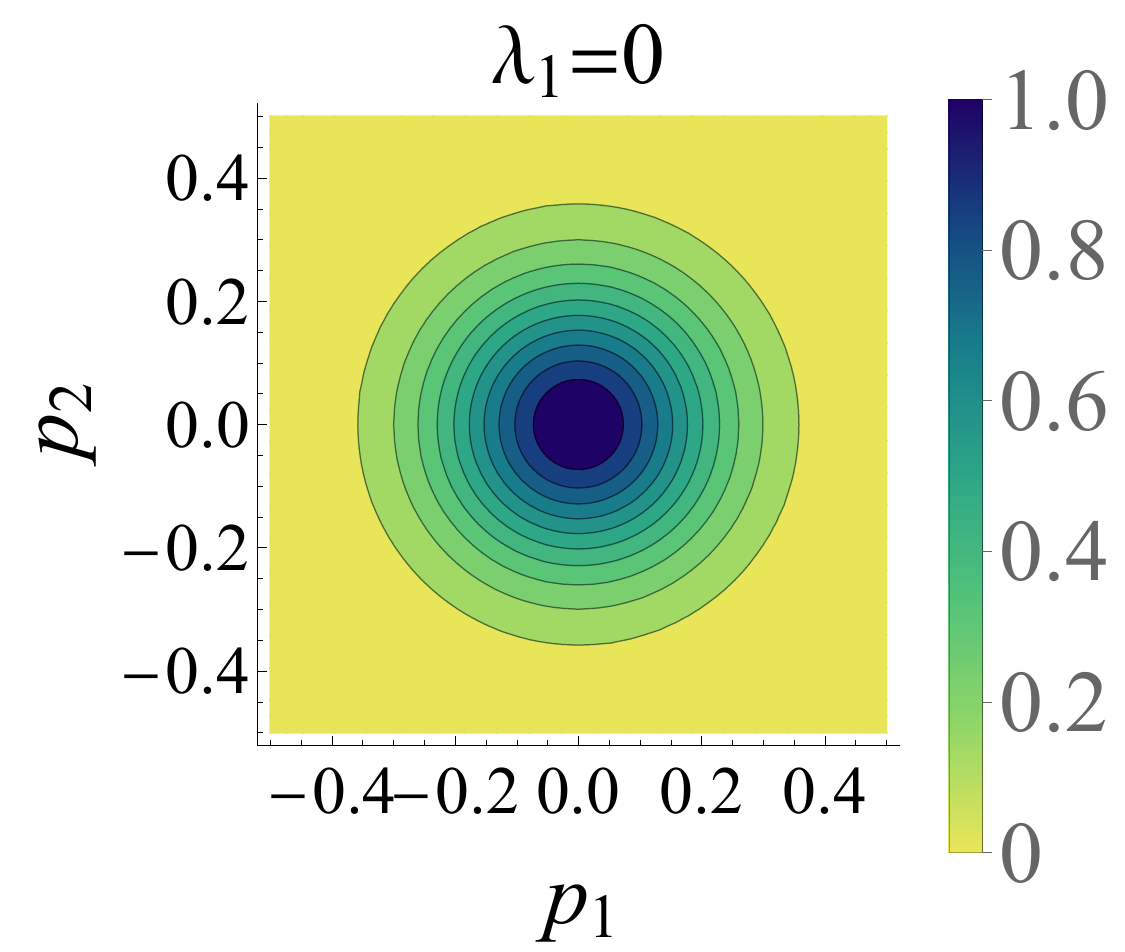}
		\includegraphics[width=4.2cm]{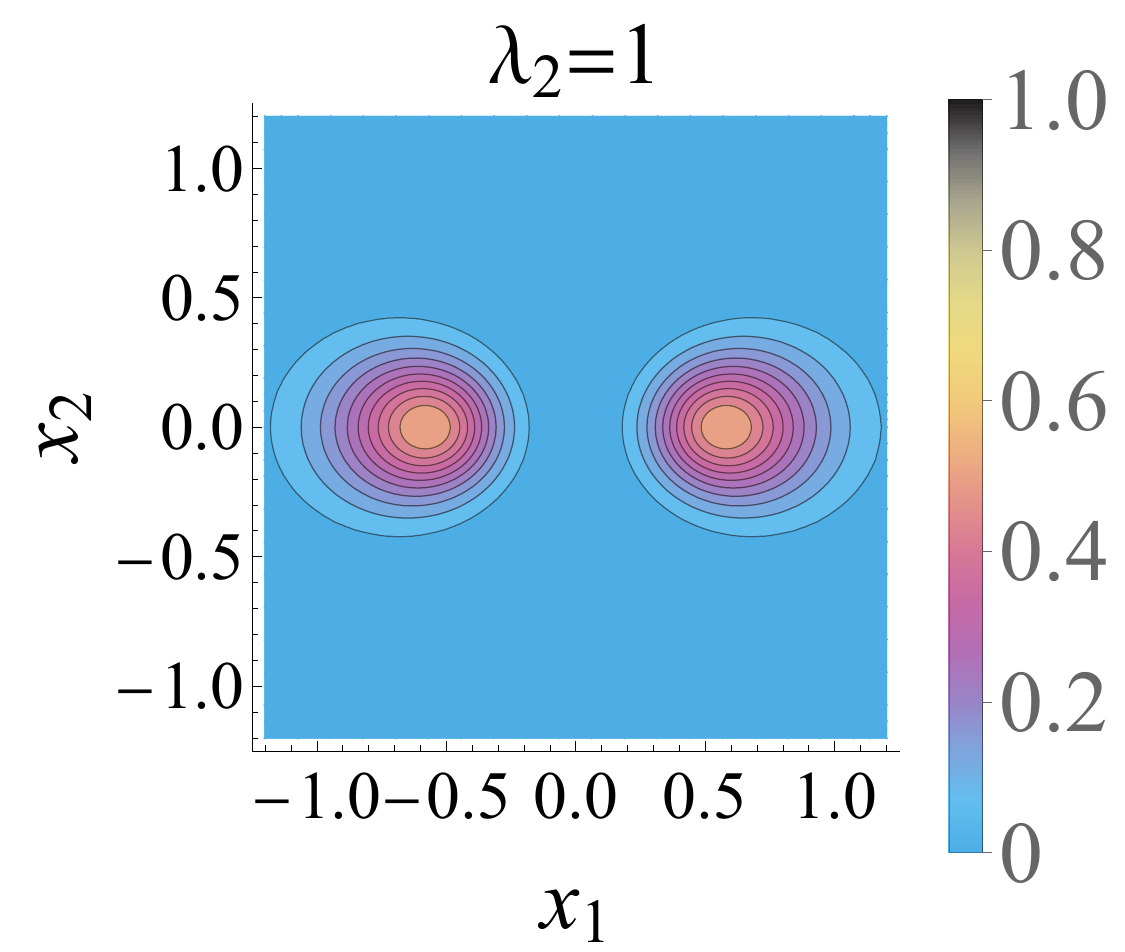}
		\includegraphics[width=4.2cm]{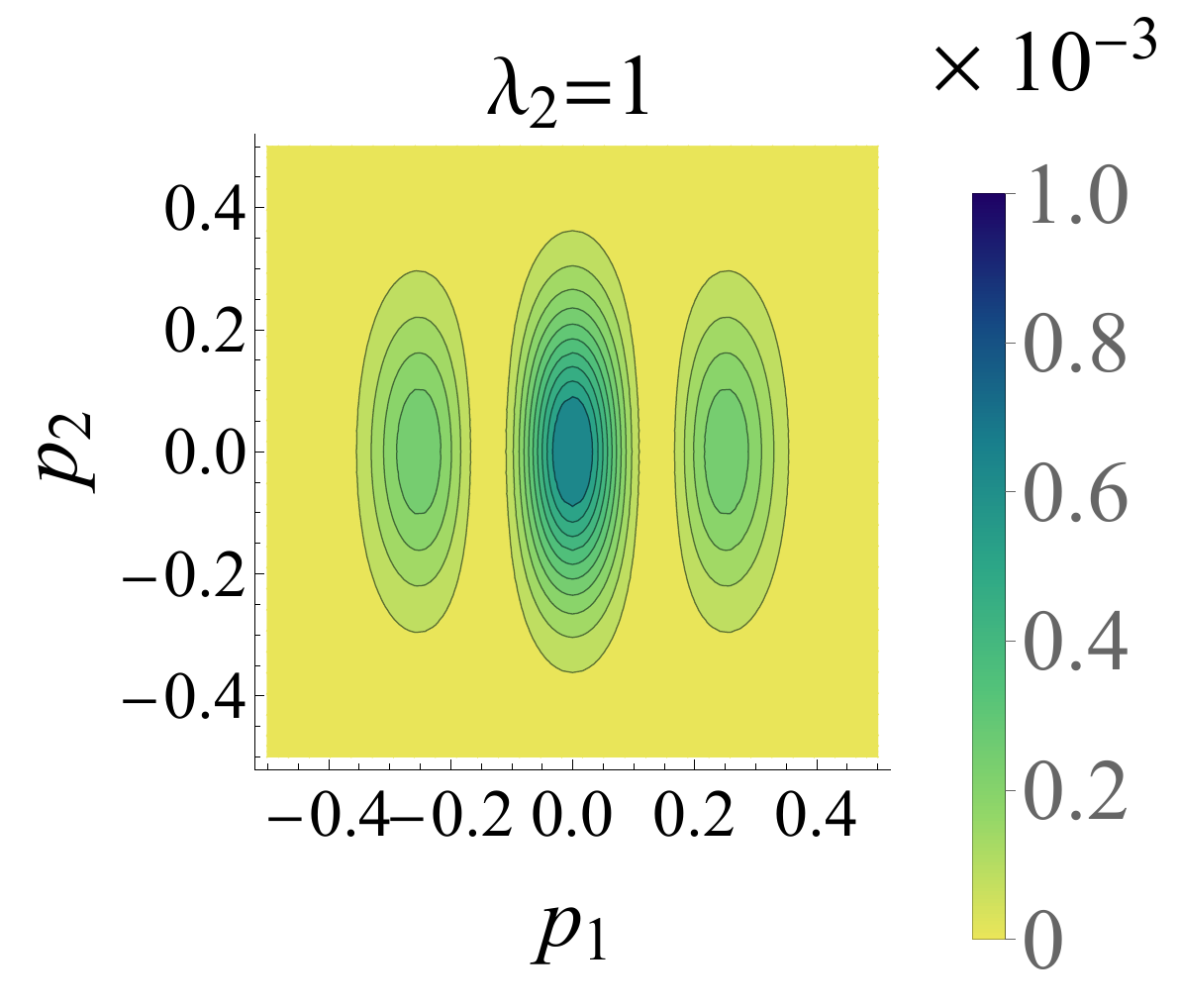}
		\includegraphics[width=4.2cm]{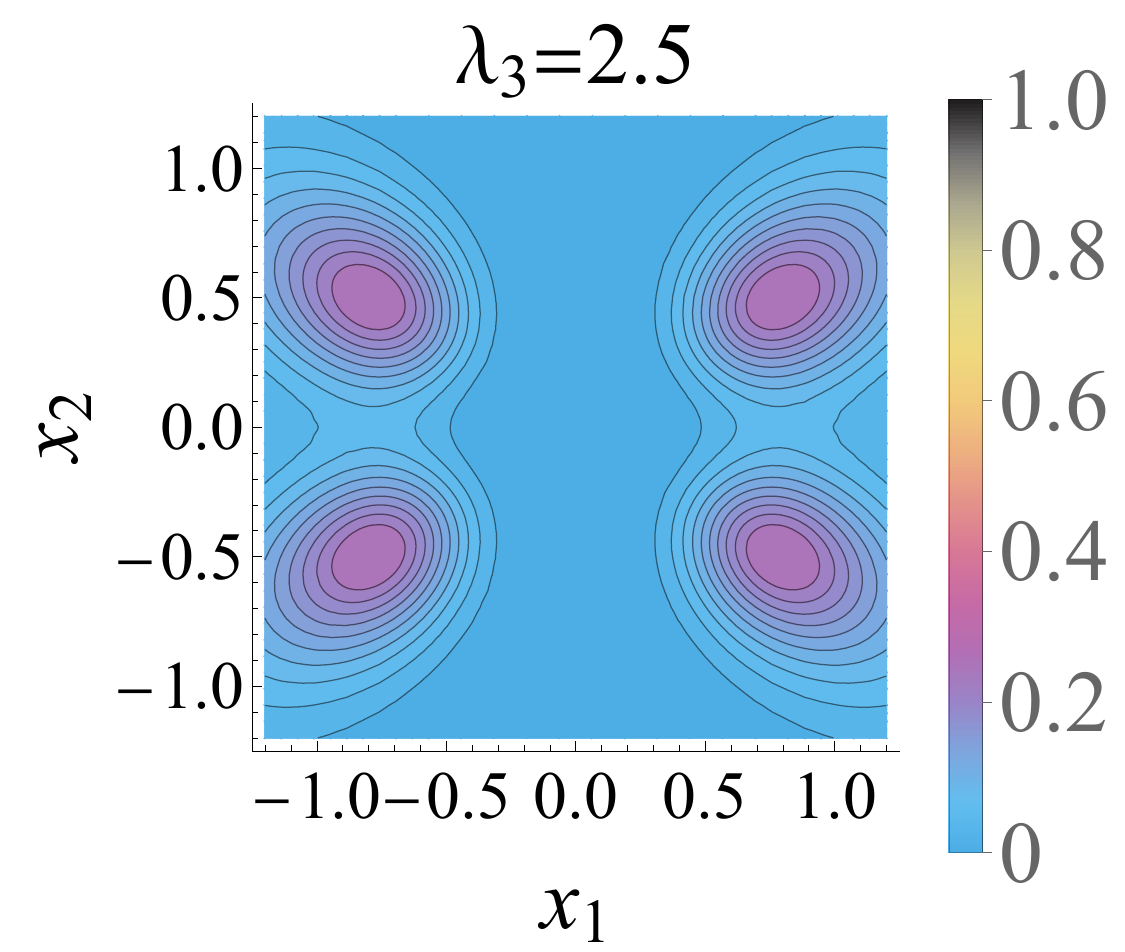}
		\includegraphics[width=4.2cm]{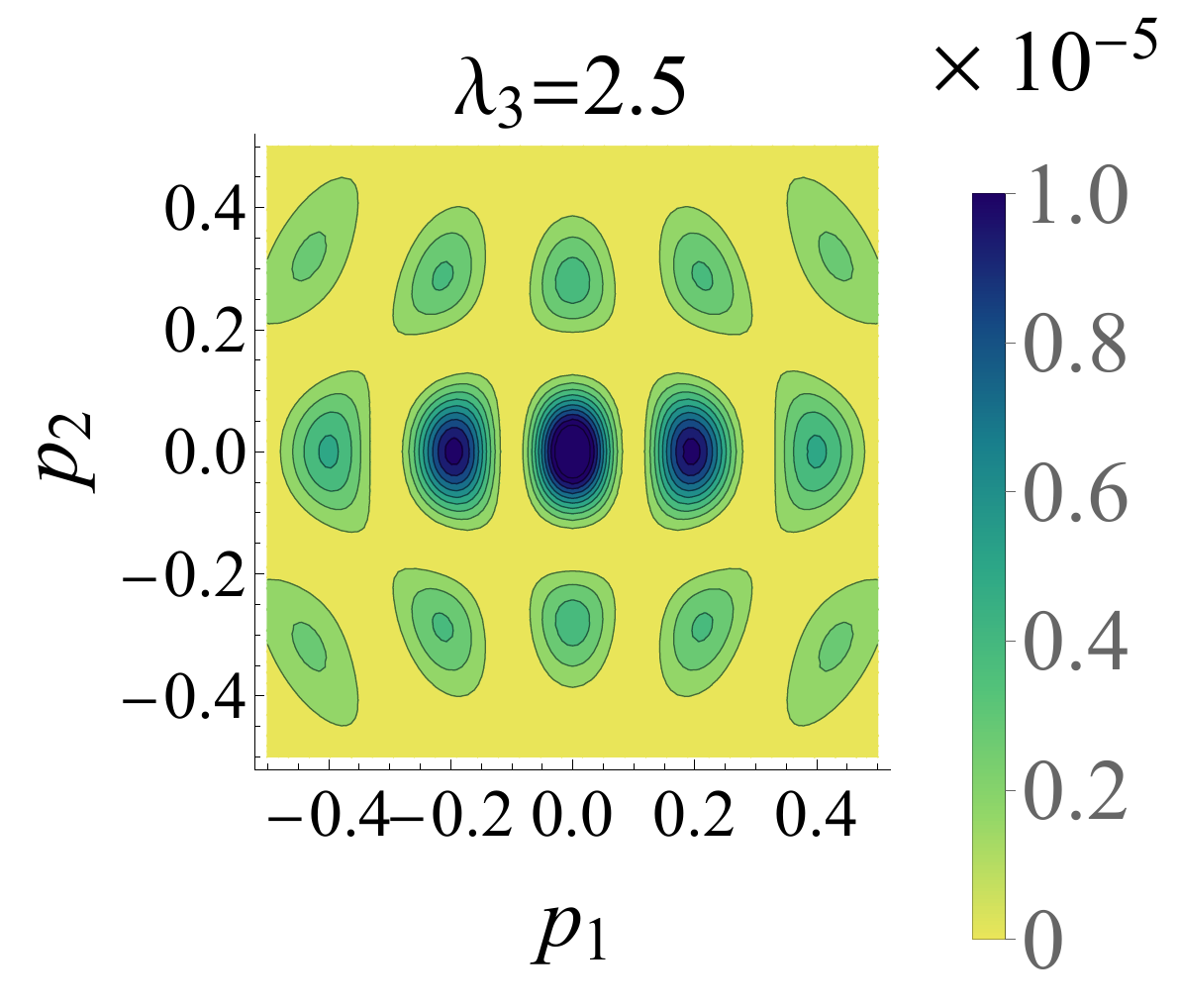}
	\end{center}
	\caption{Contour plots in phase space coordinates of the Husimi function  $Q_{|\zb^{(0)}\ra_{\mathbb{0}}}(\zb')$ of the variational ground state $|\zb^{(0)}\ra_{\mathbb{0}}$  of the LMG U(3) model (\ref{critalphabeta},\ref{vargroundNorm}), for $N=20$ particles and three  different values of the control parameter $\lambda$ ($\epsilon$ units) inside the three phases I, II and III. The left and right columns correspond to ``position"  $x_{1,2}=\text{Re}(z'_{1,2})$ and ``momentum"  $p_{1,2}=\text{Im}(z'_{1,2})$ coordinates, respectively. 
	}
	\label{Figure:HusimiFuncSym3CAT_Variational_LMG}
\end{figure}

In Figure \ref{Figure:HusimiFuncSym3CAT_Variational_LMG} we make contour plots of  the variational Husimi function in position (left panel) and momentum (right panel) spaces for three different characteristic values $\lambda_1,\lambda_2,\lambda_3$, 
\be\lambda_1=0 < \lambda^{(0)}_{\mathrm{I}\leftrightarrow\mathrm{II}}<\lambda_2=1 < \lambda^{(0)}_{\mathrm{II}\leftrightarrow\mathrm{III}}<\lambda_3=2.5,
\ee
of the control parameter $\lambda$ inside each phase for $N=20$ particles. Contour plots of $Q_{|\zb^{(0)}\ra_{\mathbb{0}}}(\zb)$ in position space give a clear visual explanation of the delocalization of the ground state in phase space as we move from phase I to phases II and III. 
Indeed, the Husimi function is composed of a single lump/hump/packet in phase I, which coincides with $2^k=1$ for $k=0$, the number of non-zero components of $\zb^{(0)}=(z_{1+}^{(0)},z_{2+}^{(0)})$ according to \eqref{critalphabeta}; similarly, we have $2^k=2$ and $2^k=4$ lumps in phases II and III for $k=1$ and $k=2$ non-zero components of $\zb^{(0)}$, respectively. The behavior of the Husimi function in momentum space is a little bit more subtle, as it entails some modulations which, in the large $N$ limit, correspond to a (Gaussian-like) packet  modulated by a
cosine function which oscillates rapidly for high $N$ mainly in phase III (see  \cite{PhysRevA.85.053831} for a similar behavior in the Dicke model in the superradiant phase).

\begin{figure}[h]
	\begin{center}
		\includegraphics[width=0.48\textwidth]{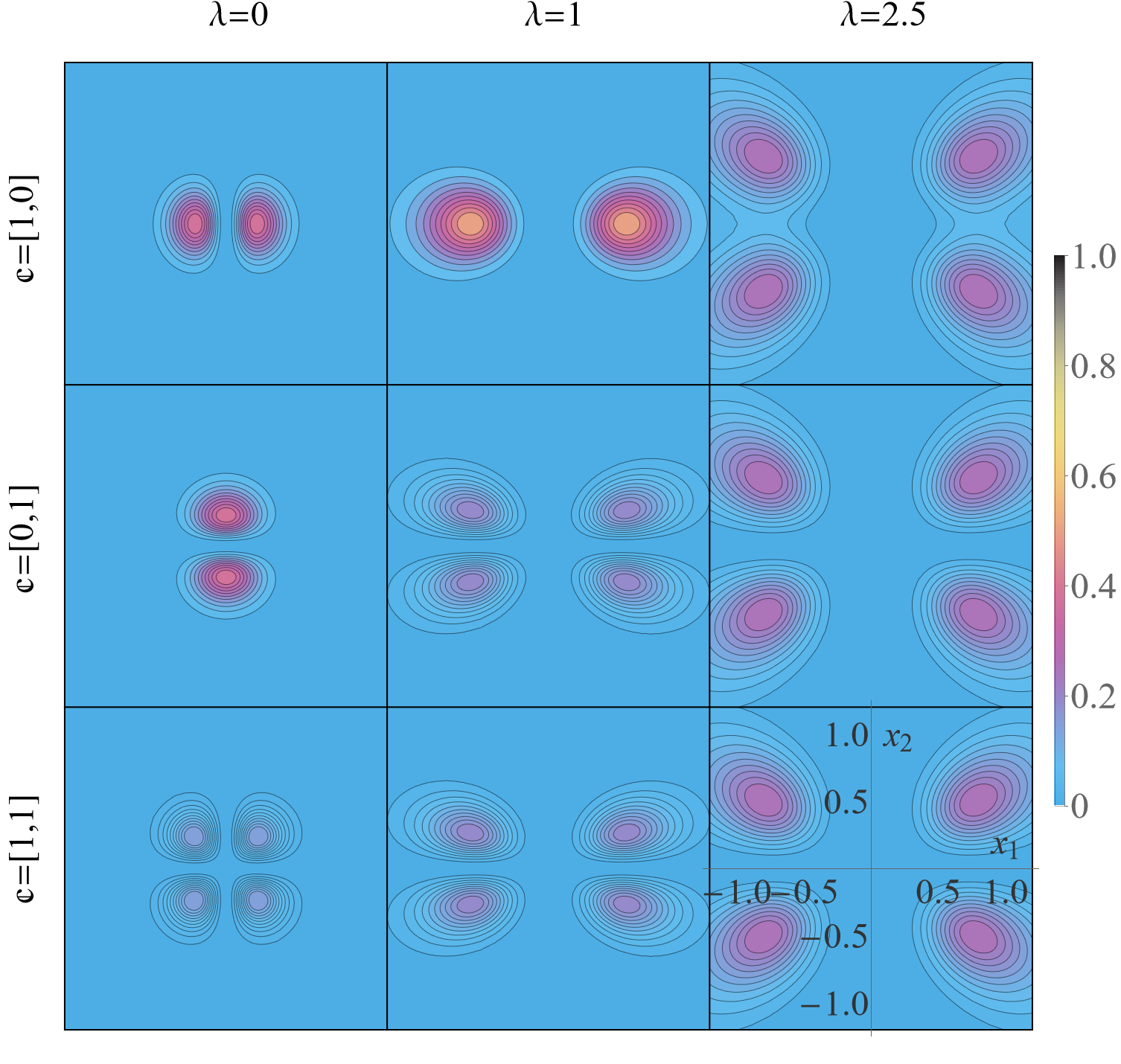}
	\end{center}
	\caption{Contour plots in phase space position coordinates $x_{1,2}=\text{Re}(z'_{1,2})$ of the Husimi function $Q_{|\zb^{(0)}\ra_{\mathbbm{c}}}(\zb')$, where $|\zb^{(0)}\ra_{\mathbbm{c}}$  are the variational excited states \eqref{varES} of the LMG U(3) model \eqref{critalphabeta}. We have chosen $N=20$ particles and three  different values of the control parameter $\lambda$ ($\epsilon$ units) inside the three phases I, II and III (columns from left to right). Each row in the plot represents a variational excited state of definite $\mathbbm{c}$-parity.
	}
	\label{Figure:HusimiFunc_c-3CATs_Variational_LMG}
\end{figure}

Additionally, in Figure \ref{Figure:HusimiFunc_c-3CATs_Variational_LMG} we study the Husimi function $Q_{|\zb^{(0)}\ra_{\mathbbm{c}}}(\zb')$ of variational excited states $i=1,3,5$ of the LMG U(3) model (already defined in eqs.(\ref{critalphabeta},\ref{varES}) and classified in Figure \ref{Figure:Energies}). We shall restrict the plot and discussion to position coordinates $x_{1,2}=\text{Re}(z'_{1,2})$ for convenience. It is interesting that, in the phase I at $\lambda=0$ (left column in Figure \ref{Figure:HusimiFunc_c-3CATs_Variational_LMG}), the variational ES Husimi functions have more than a single hump, which was not the case of the GS in the Figure \ref{Figure:HusimiFuncSym3CAT_Variational_LMG}. This is because the variational ES $|\zb^{(0)}\ra_{\mathbbm{c}}$ preserve their parity $\mathbbm{c}\neq\mathbb{0}$ even when $\zb^{(0)}\xrightarrow{\lambda\to 0}(0,0)$. Actually, this limit was already given in the eq. \eqref{3CAT_limits}. For instance, the variational first ES $\mathbbm{c}=[1,0]$ (top row in the Figure \ref{Figure:HusimiFunc_c-3CATs_Variational_LMG}) transforms into a Fock state $|\zb^{(0)}\ra_{[1,0]}^{(N)}\xrightarrow{\lambda\to 0}|{\scriptstyle n_0=N-1,\,n_1=1,\,n_2=0}\ra$. Having only one particle in level 1, $n_1=1$, implies odd-parity in $x_1=\text{Re}(z_1')$ when plotting $Q_{\zb^{(0)}_{[1,0]}}(\zb')$ (check the eqs. \eqref{coefCS} and \eqref{HusimiDef}). Therefore, the variational first ES cannot be 0 at $x_1=x_2=0$ and has two humps along the $x_1$-axis direction (top left panel in Figure \ref{Figure:HusimiFunc_c-3CATs_Variational_LMG}). The variational third ES $\mathbbm{c}=[1,0]$ (middle row) has a similar behavior at $\lambda=0$ but along the $x_2$-axis, $|\zb^{(0)}\ra_{[0,1]}^{(N)}\xrightarrow{\lambda\to 0}|{\scriptstyle n_0=N-1,\,n_1=0,\,n_2=1}\ra$. The fifth ES $\mathbbm{c}=[1,1]$ (bottom row) has double odd-parity in the axis $x_1$ and $x_2$ and presents four humps, $|\zb^{(0)}\ra_{[0,1]}^{(N)}\xrightarrow{\lambda\to 0}|{\scriptstyle n_0=N-2,\,n_1=1,\,n_2=1}\ra$. In the phase II at $\lambda=1$ (middle column in Figure \ref{Figure:HusimiFunc_c-3CATs_Variational_LMG}), all the Husimi functions of the variational ES have symmetric humps along the $x_1$-axis as the GS did in the Figure \ref{Figure:HusimiFuncSym3CAT_Variational_LMG}. However, the third and fifth ES also display symmetric humps along the $x_2$-axis, as both have $c_2=1$ in $\mathbbm{c}$. Finally, in the phase III at $\lambda=2.5$ (right column in Figure \ref{Figure:HusimiFunc_c-3CATs_Variational_LMG}), the ESs have four humps as the GS, demonstrating the degeneration already showed in the Figure \ref{Figure:Energies} at $\lambda\gg 1$. This result agrees with the eq.\eqref{HusimiDCATLimit} in Appendix \ref{app2} (number of terms in the sum $\sum_{\mathbbm{b}\in\{0,1\}^2}$), but for relatively large finite ($N=20$) number of particles.

As a general rule, we propose that the number of humps (in the phase space coordinates $\zb'$) of a $\mathbbm{c}$-DCAT Husimi function is
\begin{equation}\label{numHumps}
	\#_{\text{humps}}\left(Q_{|\zb\ra_{\mathbbm{c}}}(\zb')\right)=2^{\|\zb\|_0+\|\mathbbm{c}_L\|_0}\quad\forall N>>1\,,
\end{equation}
where $K=\{j_1,\ldots,j_k\}$ and $L=\{i_1,\ldots,i_l\}$ are the set of indexes of the non-zero and zero coordinates in $\zb$ respectively, and $k=\|\zb\|_0$ and $\|\mathbbm{c}_L\|_0$ are the number of non-zero components in $\zb$ and $\mathbbm{c}_L$ respectively (see the eqs.(\ref{symDCAT_limits2},\ref{cDCATs_limits2},\ref{mumomentcDCATlimit2}) to revisit the notation). The proof of this proposition is based on the thermodynamic limit of $Q_{|\zb\ra_{\mathbbm{c}}}(\zb')$ and its $\nu$-moments \eqref{mumomentcDCATlimit2}. The number of humps in the expression above cannot be greater than $2^{D-1}$, as $\|\zb\|_0+\|\mathbbm{c}_L\|_0\leq D-1$, where $\|\zb\|_0=k\leq D-1$ and $\|\mathbbm{c}_L\|_0\leq l=D-1-k$ by construction. For instance, in the case $D=3$, we have a maximum of $2^2=4$ humps, like in the Figures \ref{Figure:HusimiFuncSym3CAT_Variational_LMG} and \ref{Figure:HusimiFunc_c-3CATs_Variational_LMG}. If we focus on the eq.\eqref{cDCATs_limits2}, we realize that $2^{\|\zb\|_0}$ is the number of humps of the reduced $\mathbbm{c}_K$-DCAT $|(\zb_K,\zb_L=\bm{0}_L)\ra_{\mathbbm{c}_K}^{(N-\|\mathbbm{c}_L\|_0)}$ in the thermodynamic limit, while the Fock state $|{\scriptstyle \vec{n}_K=\vec{0}_K,\vec{n}_L=\mathbbm{c}_L}\ra$ has $2^{\|\mathbbm{c}_L\|_0}$ humps by construction. The reduced $\mathbbm{c}_K$-DCAT coordinates $\zb_K$ are non-zero by definition, so $\|\zb\|_0=k$ and we obtain the maximum number of humps $2^k$ allowed in a reduced phase space with $k$ coordinates. In the case of the fully even DCAT, $\mathbbm{c}=\mathbb{0}$ and $\|\mathbbm{c}_L\|_0=0$, we recover the results of the Figure \ref{Figure:HusimiFuncSym3CAT_Variational_LMG} and the equation \eqref{HusimiDCATLimitRed}. We shall also highlight that $2^{\|\zb\|_0+\|\mathbbm{c}_L\|_0}$ is also the rank of the $M$-particle reduced density matrix of a $\mathbbm{c}$-DCAT, as it is shown in \cite{DcatDecompositionArxiv}. The connection of the two concepts is subject to further investigation.

\begin{figure}[h]
	\begin{center}
		\includegraphics[width=0.48\textwidth]{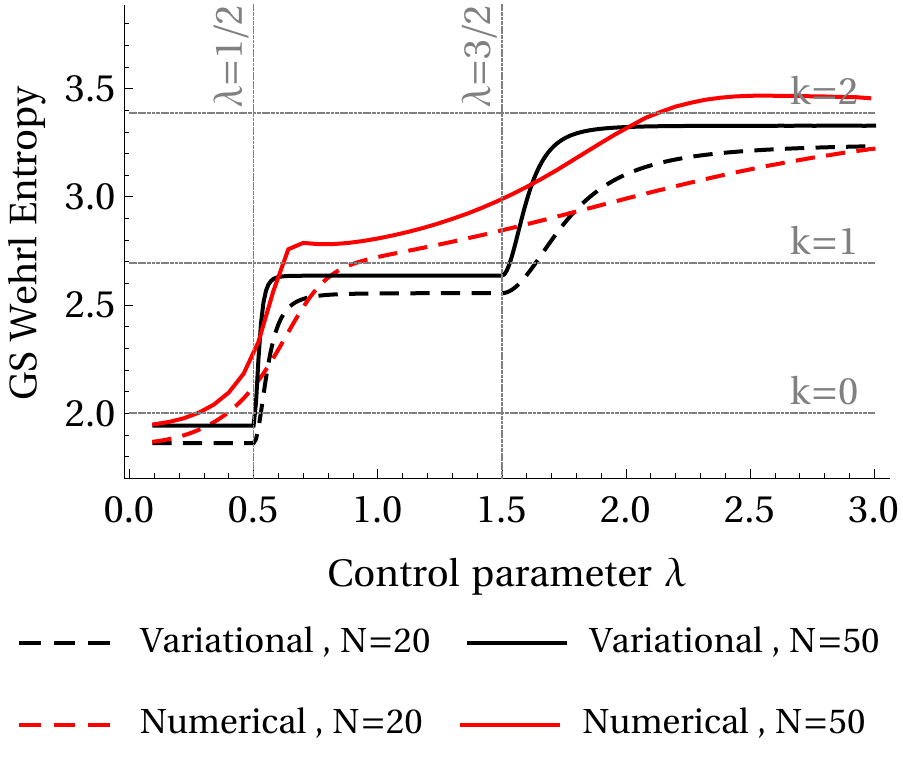}
	\end{center}
	\caption{Wehrl entropy of the {variational} $|\zb^{(0)}\ra_\mathbb{0}$ and numerical $|\psi_0\ra$ ground state of the LMG U(3) model for $N=20$ and $50$ particles. The gray dashed vertical lines represent the quantum critical points at $\lambda^{(0)}_{\mathrm{I}\leftrightarrow\mathrm{II}}=1/2$ and 
		$\lambda^{(0)}_{\mathrm{II}\leftrightarrow\mathrm{III}}=3/2$ (in $\epsilon$ units). The gray dashed horizontal lines are the $N\to\infty$ limits of the Wehrl entropy of the $\mathbb{0}$-3CAT $|\zb^{(0)}\ra_\mathbb{0}$ \eqref{WehrlEntDCATlimit2}, with $k$ humps (the number of non-zero coordinates in $\zb^{(0)}(\lambda)$). 
		}
	\label{Figure:WehrlEntropySym3CAT_Variational_LMG}
\end{figure}

The delocalization (area) of the Husimi function in phase space, which is perceived in Figures \ref{Figure:HusimiFuncSym3CAT_Variational_LMG} and \ref{Figure:HusimiFunc_c-3CATs_Variational_LMG} across the different phase transitions, can be quantified by using the Wehrl entropy \eqref{WehrlIntegral}. In Figure \ref{Figure:WehrlEntropySym3CAT_Variational_LMG}, we present the Wehrl entropy of the variational (black curves) and numerical (red curves)  ground state (GS) of the LMG U(3) model, as a function of the control parameter $\lambda$ for different values of $N$. The entropy suddenly grows  around the quantum critical points $\lambda^{(0)}_{\mathrm{I}\leftrightarrow\mathrm{II}}=1/2$ and 
$\lambda^{(0)}_{\mathrm{II}\leftrightarrow\mathrm{III}}=3/2$, which are represented with vertical dashed lines. This effect is more abrupt with increasing $N$. In addition, the values of the entropy in each phase tend to the thermodynamic limit of the 3CAT entropy \eqref{WehrlEntDCATlimit2}, with different number $k$ of non-zero components in $\zb$. In particular for $D=3$, this limit is $2+k\log(2)$ with $k=0,1,$ and $2$ in the phases I, II, and III respectively, which corresponds to the gray dashed horizontal lines in the Figure \ref{Figure:WehrlEntropySym3CAT_Variational_LMG}. When there is a QPT in the LMG U(3) model, the GS Husimi function in the position space (left column in Figure \ref{Figure:HusimiFuncSym3CAT_Variational_LMG}) splits into two identical subpackets with negligible overlap, so the Wehrl entropy experiences an increment of $\ln(2)$ (see \cite{real2013} for a similar result in the case of the Dicke model of superradiance). This delocalization effect happens twice from the phase I to the III, hence the $2^2$ subpackets of the Husimi function in the phase III and the total growth of $2\log(2)$ in the Wehrl entropy. 


The ``Numerical'' red curves in the Figure \ref{Figure:WehrlEntropySym3CAT_Variational_LMG} refer to the ground state obtained by numerical diagonalization of the Hamiltonian \eqref{hamU3}. The eigenvectors are calculated in the Fock basis \eqref{psisym}, introduced in the Husimi function equation \eqref{HusimiDef}, and then, the Wehrl function \eqref{WehrlIntegral} is numerically integrated. The change of entropy in the numerical (exact) case (red curves)  is less abrupt than in the variational one (black curves) around the quantum critical points for a given number of particles $N$, although it becomes steeper and steeper as $N$ increases.

\begin{figure}[h]
	\begin{center}
		\includegraphics[width=0.48\textwidth]{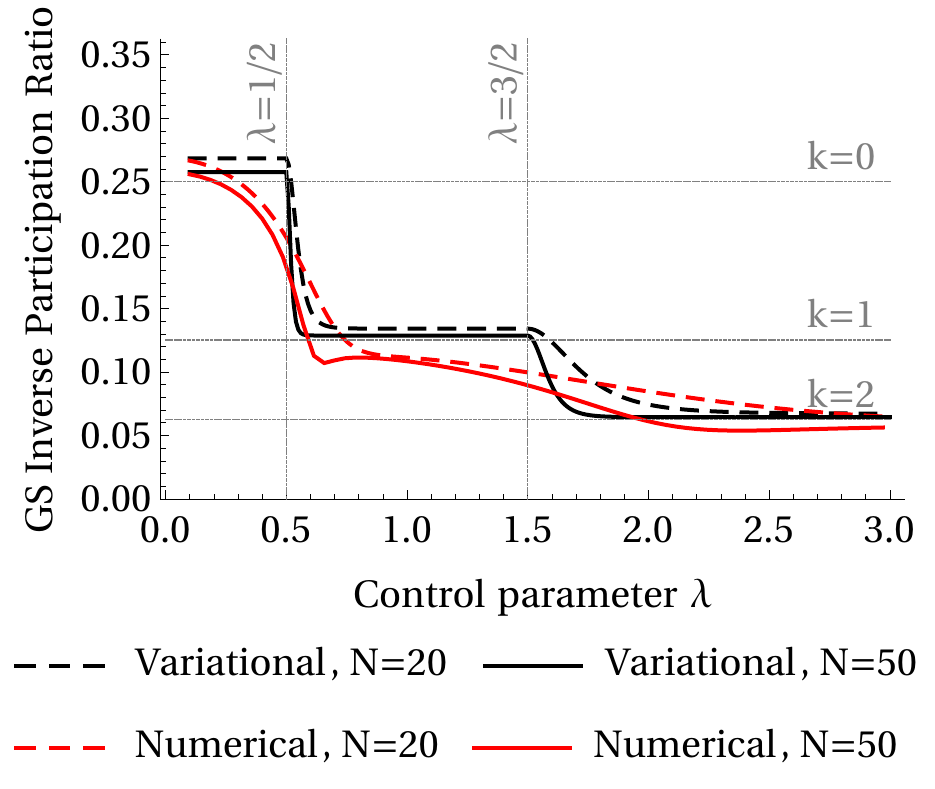}
	\end{center}
	\caption{Inverse Participation Ratio (IPR) of the variational $|\zb^{(0)}\ra_\mathbb{0}$ and numerical $|\psi_0\ra$ ground state of the LMG U(3) model for $N=20$ and $50$ particles. The gray dashed vertical lines represent the critical points at $\lambda^{(0)}_{\mathrm{I}\leftrightarrow\mathrm{II}}=1/2$ and 
		$\lambda^{(0)}_{\mathrm{II}\leftrightarrow\mathrm{III}}=3/2$ (in $\epsilon$ units). The gray dashed horizontal lines are the $N\to\infty$ limit of the IPR of the $\mathbb{0}$-3CAT according to \eqref{mumomentDCATlimit2} for $\nu=2$ and $D=3$, that is, 
		$\lim_{N\to\infty}M_2(|\zb^{(0)}\ra_\mathbb{0})=2^{-k-2}=\{\frac{1}{4},\frac{1}{8},\frac{1}{16}\}$, for $k=0,1,2$ the number of non-zero components in $\zb^{(0)}(\lambda)$.}
	\label{Figure:IPRSym3CAT_NumNVar_LMG}
\end{figure}

Equivalently, one can  also measure  the localization of the ground state in phase space with the IPR or the Husimi second moment \eqref{momentsNu}. This quantity is usually easier (and faster) to calculate than the Wehrl entropy. 
That is why it is more common to focus on the IPR when studying localization \cite{Berke2022,Giannini2019,Giannini2022}. 
The IPR of the ground state attains the thermodynamic limit value presented in the equation \eqref{mumomentDCATlimit2} for $\nu=2$ and $k=0,1,2$. Variational calculations provide sharper results than the numerical ones. For large values of the control parameter $\lambda$, the ground state behaves as a 3CAT which is less localized than the DSCS in  phase space (check out Husimi function in Figure \ref{Figure:HusimiFuncSym3CAT_Variational_LMG}), and therefore, Figure \ref{Figure:IPRSym3CAT_NumNVar_LMG} shows a decrease of the IPR when  increasing $\lambda$.

\begin{figure}[h]
	\begin{center}
		\includegraphics[width=0.48\textwidth]{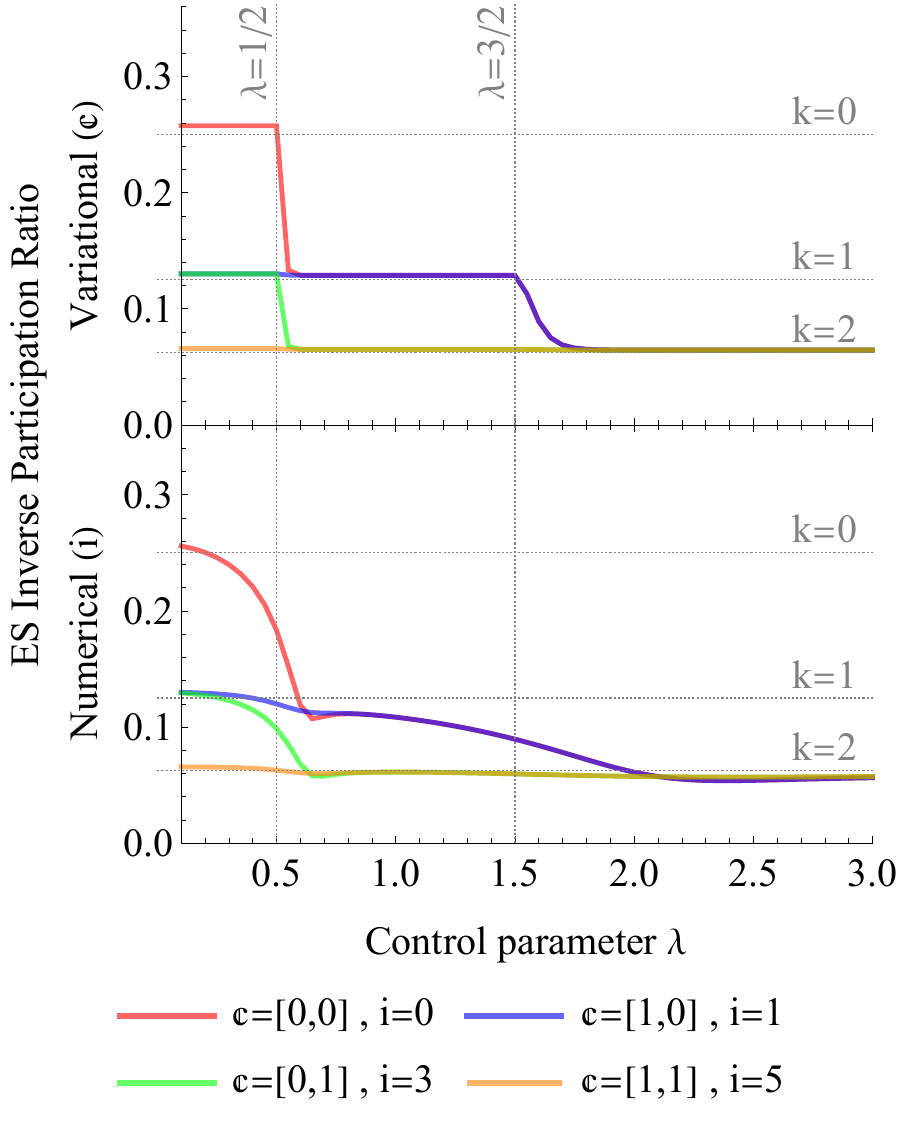}
	\end{center}
	\caption{Inverse Participation Ratio (IPR) of the variational $|\zb^{(0)}\ra_\mathbbm{c}$ (top panel) and numerical $|\psi_i\ra$ (bottom panel) excited states of the LMG U(3) model for $N=50$ particles. The gray dashed vertical lines represent the critical points at $\lambda^{(0)}_{\mathrm{I}\leftrightarrow\mathrm{II}}=1/2$ and 
		$\lambda^{(0)}_{\mathrm{II}\leftrightarrow\mathrm{III}}=3/2$ (in $\epsilon$ units). The gray dashed horizontal lines are the $N\to\infty$ limit of the IPR of the $\mathbbm{c}$-3CAT according to \eqref{mumomentcDCATlimit2} for $\nu=2$ and $D=3$, that is, 
		$\lim_{N\to\infty}M_2(|\zb\ra_\mathbbm{c}^{(0)})=2^{-k-\|\mathbbm{c}_L\|_0-2}=\{\frac{1}{4},\frac{1}{8},\frac{1}{16}\}$, for $k+\|\mathbbm{c}_L\|_0=0,1,2$ the possible number of humps of $Q_{|\zb^{(0)}\ra_\mathbbm{c}}(\zb')$ for $N>>1$ \eqref{numHumps}.}
	\label{Figure:IPRc3CATs_NumNVar_LMG}
\end{figure}

As the IPR numerical computation is faster than the Wehrl entropy one, it is also feasible to reproduce the Figure \ref{Figure:IPRSym3CAT_NumNVar_LMG} but for the ESs of the LMG U(3) model. In particular, Figure \ref{Figure:IPRc3CATs_NumNVar_LMG} shows the IPR of the numerical ESs $|\psi_i\ra$, $i=0,1,3,5$, and its associated variational ESs $|\zb^{(0)}\ra_\mathbbm{c}$ regarding the equation \eqref{varES}, where we have used $N=20$ particles and the color code is the same as in the energy spectrum in Figure \ref{Figure:Energies}. In the top panel, the variational ESs approximate faster to the gray dashed horizontal lines (eq.\eqref{mumomentcDCATlimit2} for $\nu=2$ and $k+\|\mathbbm{c}_L\|_0=0,1,2$) than the numerical ones in the bottom panel, as it happened in the Figure \ref{Figure:IPRSym3CAT_NumNVar_LMG} for the GS. The three different phases of the Figure \ref{Figure:IPRc3CATs_NumNVar_LMG} are delimited by the gray dashed vertical lines, so that in the phases I, II and III there are $k=\|\zb^{(0)}(\lambda)\|_0=0,1,2$ non-zero coordinates in $\zb^{(0)}(\lambda)$ (see eq.\eqref{critalphabeta}). Therefore, the IPR of the ESs reaches the gray dashed horizontal lines according the number of humps displayed in the Figures \ref{Figure:HusimiFuncSym3CAT_Variational_LMG} and \ref{Figure:HusimiFunc_c-3CATs_Variational_LMG}, which depends on $k$ and $\|\mathbbm{c}_L\|_0$ as $2^{k+\|\mathbbm{c}_L\|_0}$ (see eq.\eqref{numHumps}). That is, for example, for $\mathbbm{c}=[1,0]$ or $i=1$ (blue line), the ES has two ($k=0$, $\|\mathbbm{c}_L\|_0=1$), two ($k=1$, $\|\mathbbm{c}_L\|_0=0$), and four ($k=2$, $\|\mathbbm{c}_L\|_0=0$) humps in the three respective phases of the Figure \ref{Figure:HusimiFunc_c-3CATs_Variational_LMG} (top row); hence, it attains the values $k+\|\mathbbm{c}_L\|_0=1,1,2$  marked by gray dashed horizontal lines in each phase of the Figure \ref{Figure:IPRc3CATs_NumNVar_LMG}, respectively. This result is in agreement with the general expression in eq. \eqref{mumomentcDCATlimit2} for the thermodynamic limit of the $\mathbbm{c}$-DCAT Husimi moments  for $\nu=2$ and  $D=3$.

\section{Conclusions}\label{conclusec}

	
	The concept of Husimi function in the canonical phase space is extended to the complex projective space $\mathbb{C}P^{D-1}=\text{U}(D)/[\text{U}(1)\times\text{U}(D-1)]$ using $\text{U}(D)$-spin coherent states  (DSCSs for short) for symmetric multi-quDit systems. The $\nu$-moments of the Husimi function and some localization measures in phase space such as the Inverse Participation Ratio  and the Wehrl entropy are accordingly extended with a proper integration (Harr) measure. We prove that the Lieb conjecture is fulfilled for the DSCSs in the eq.\eqref{mumomentD3} and the Appendix \ref{app1}. The parity $\mathbb{Z}_2^{D-1}\ni \mathbbm{c}$  adaptations of DSCSs (called $\mathbbm{c}$-DCAT states) turn out to be less localized than the DSCSs, which exhibit maximum localization (minimum area in phase space) according to Lieb's conjecture. This becomes clear when we calculate the thermodynamic limit of the Husimi function $\nu$-th moments and Wehrl entropy for  DSCSs and  $\mathbbm{c}$-DCAT states. 
	
	The previous study of the LMG U(3) ground state \cite{nuestroPRE} is then extended to the first excited states, which turn out to be modeled by $\mathbbm{c}$-3CATs of different parities, as Figures \ref{Figure:Energies} and \ref{Figure:Fidelityfig1} show. In particular, we compare the numerical eigenstates of the LMG U(3) model (for finite $N$) to different variational $\mathbbm{c}$-3CATs states via fidelity \eqref{fidEq}, where the variational states are evaluated at the critical points $\zb^{(0)}=(z_{1\pm}^{(0)},z_{2\pm}^{(0)})$ which minimize the LMG U(3) energy surface in the thermodynamic limit \eqref{critalphabeta}. The variational $\mathbbm{c}$-3CAT states turn out to be fairly faithful to the low-lying excited Hamiltonian eigenstates except in the vicinity of the critical points  $\lambda^{(0)}_{\mathrm{I}\leftrightarrow\mathrm{II}}=\epsilon/2$ and $\lambda^{(0)}_{\mathrm{II}\leftrightarrow\mathrm{III}}=3\epsilon/2$ separating quantum phases I, II and III. We believe this is a consequence of the growth of quantum fluctuations at the critical points. However, this fidelity can be improved by maximizing the corresponding overlap in the complex projective phase space $\mathbb{C}P^2\ni\zb$, as we display in the Figures \ref{Figure:MaximizationOverlapZcrit} and \ref{Figure:MaximizationOverlap}. 
	
	The fact that the minimization of the energy surface in the thermodynamic limit provides critical vectors $\zb^{(0)}$ with some zero components in certain phases, makes it necessary to revise the $\mathbb{Z}_2^{D-1}$-parity adaptation $|\zb^{(0)}\ra_\mathbbm{c}$ of  $|\zb^{(0)}\ra_\mathbbm{c}$ and to resolve some ``$0/0$'' indeterminacies. In the case when $\zb$ has  $l=D-1-k$ null coordinates, 
	the corresponding $\mathbbm{c}$-DCAT  $|\zb^{(0)}\ra_\mathbbm{c}$ reduces to $\mathbbm{c}_K$-DCATs with lower  $\mathbb{Z}_2^{k}$-parity \eqref{cDCATs_limits2}. This result permeates in the majority of magnitudes (Husimi function, its moments, etc.) calculated in this work.

	The QPTs of the LMG U(3) model  are visualized  in the phase space $\mathbb{C}P^2\ni\zb'$ across the phase diagram via the Husimi function $Q_\mathbb{0}(\zb')$ of the variational ground state $|\zb^{(0)}\ra_\mathbb{0}$. We draw contour plots of the Husimi function in  ``position space'' $(x_1,x_2)=\text{Re}(\zb')$ and in ``momentum space'' $(p_1,p_2)=\text{Im}(\zb')$ in Figure \ref{Figure:HusimiFuncSym3CAT_Variational_LMG}). In position space, the variational GS Husimi function $Q_\mathbb{0}(\zb')$ displays several humps depending on the number of non-zero coordinates of $\zb^{(0)}(\lambda)$,  which changes in the different quantum phases I, II and III. A similar reasoning is followed in Figure \ref{Figure:HusimiFunc_c-3CATs_Variational_LMG} with the Husimi function of the other variational $\mathbbm{c}$-3CAT states $|\zb^{(0)}\ra_\mathbbm{c}$ mimicking low-lying Hamiltonian eigenstates with parity $\mathbbm{c}$. We propose a general   expression \eqref{numHumps} for the number of humps (in position phase) of the Husimi function of general $\mathbbm{c}$-3CATs $|\zb\ra_\mathbbm{c}$, depending on the number of zero components of $\zb$ and the  parity $\mathbbm{c}$. This number also appears in the thermodynamic limit of the $\mathbbm{c}$-DCAT Husimi moments \eqref{mumomentcDCATlimit2} and in the rank of the $M$-particle reduced density matrix of a $\mathbbm{c}$-DCAT \cite{DcatDecompositionArxiv}.
	
	Finally, we also characterize the QPTs via localization measures in phase space, since the Husimi fuction $Q_\mathbb{0}$ of the ground state of the LMG model suddenly suffers delocalization when passing through the quantum critical $\lambda^{(0)}$ points, as shwon in Figures of  Wehrl entropy  \ref{Figure:WehrlEntropySym3CAT_Variational_LMG} and IPR  \ref{Figure:IPRSym3CAT_NumNVar_LMG} of $Q_\mathbb{0}$ as a function of the control parameter  $\lambda$. More localization implies less Wehrl entropy (less area) and more IPR. This effect is more abrupt for the variational ground state than for the numerical one, and gets sharper and sharper when increasing $N$, approaching to the limits proposed in the Section \ref{Husimisec} and proved in the Appendix \ref{app2}. The same analysis is extended to the numerical excited states and variational $\mathbbm{c}$-3CATs in Figure \ref{Figure:IPRc3CATs_NumNVar_LMG}, which also experience delocalization, but only when its Husimi function number of humps changes according to the Figure \ref{Figure:HusimiFunc_c-3CATs_Variational_LMG}.

%
%
%
%
%

\section*{Acknowledgments}
We thank the support of the Spanish MICINN  through the project PGC2018-097831-B-I00 and  Junta de Andaluc\'\i a through the projects  UHU-1262561 and FQM-381. 
JG also thanks MICINN for financial support from FIS2017-84440-C2-2-P. 
AM thanks the Spanish MIU for the FPU19/06376 predoctoral fellowship. We all thank Octavio Casta\~nos for his valuable comments in the early stages of this work. 

\appendix 

\section{Reduced parity adapted U($D$)-spin coherent states}\label{App:DCATlimits}

We generalize the $z_i\to 0$ limits in \eqref{2CAT_limits} and \eqref{3CAT_limits} for a general $\mathbbm{c}$-DCAT. Firstly, in the fully even case $\mathbbm{c}=\mathbb{0}$, it is easy to check that the $\mathbb{0}$-DCAT in the equation \eqref{DCAT} turns into a reduced $\mathbb{0}_i$-DCAT,
\begin{equation}
	\lim\limits_{z_i\to 0}|\zb\ra_{\mathbb{0}}^{(N)}=|\zb_i\ra_{\mathbb{0}_i}^{(N)}=\frac{\Pi_{\mathbb{0}_i}}{\mathcal{N}(\zb_i)_{\mathbb{0}_i}}|\zb_i\ra^{(N)}\,,\label{symDCAT_limits}
\end{equation}
whose projective coordinates include $z_i=0$, $\zb_i=(z_1,\ldots,z_{i-1},0,z_{i+1},\ldots,z_{D-1})$, but its parity string $\mathbbm{c}$ does not contain $c_i=0$, i.e. $\mathbb{0}_i=[0,\stackrel{(D-2)}{\ldots},0]\in\mathbb{Z}_2^{D-2}$. That is, $\Pi_{\mathbb{0}_i}$ only acts onto the non-zero coordinates of $|\zb_i\ra$. Note that the reduced $\mathbb{0}_i$-DCAT is not a (D-1)CAT, as it is the $\mathbb{Z}_2^{D-2}$-parity adapted version of a DSCS with $z_i=0$, i.e.  $|\zb_i\ra^{(N)}=\lim_{z_i\to 0}|\zb\ra^{(N)}$. The normalization constant $\mathcal{N}(\zb_i)_{\mathbb{0}_i}$ is calculated as in \eqref{normcat} but using a reduced sum in $\mathbbm{b}_i\in\{0,1\}^{D-2}$, and with the new coordinates $\zb_i$,
\begin{equation}
	\mathcal{N}(\zb_i)_{\mathbb{0}_i}^2=2^{2-D} \frac{ \sum_{\mathbbm{b}_i\in\{0,1\}^{D-2}}  (1+\zb_i^\dag\zb_i^{\mathbbm{b}_i})^N}{(1+\zb_i^\dag\zb_i^{})^N}.\label{normcatRed}
\end{equation}

The zero limit \eqref{symDCAT_limits} can be used repeatedly for a set of $l=D-1-k$ different coordinates $\zb_L=\{z_{i_1},\ldots,z_{i_l}\}$, whose indexes are taken form the set $L=\{i_1,\ldots,i_l\}$, transforming the totally even $\mathbb{0}$-DCAT into a reduced $\mathbb{0}_K$-DCAT with a parity symmetry given by $\mathbb{Z}_2^{k}$,
\begin{equation}
	\lim\limits_{z_L\to \bm{0}_L}|\zb\ra_{\mathbb{0}}^{(N)}=|(\zb_K,\zb_L=\bm{0}_L)\ra_{\mathbb{0}_K}^{(N)}\,,\label{symDCAT_limits2}
\end{equation}
where $(\zb_K,\zb_L=\bm{0}_L)=\lim_{z_L\to \bm{0}_L}\zb$ has only $k$ non-zero coordinates $\zb_K=\{z_{j_1},\ldots,z_{j_k}\}$, whose associated parity components are $\mathbb{0}_K=[0,\stackrel{(k)}{\ldots},0]\in\mathbb{Z}_2^{k}$. That is, $K=\{j_1,\ldots,j_k\}$ is the set including all the $k$ non-zero coordinates of $\zb$. $\bm{0}_L$ denotes the $l$ coordinates $(0,\stackrel{l}{\ldots},0)$. The existence and uniqueness of the multiple limit \eqref{symDCAT_limits2} can be derived using hyperspherical coordinates with the moduli of $|z_i|$. The norm of the reduced $\mathbb{0}_K$-DCAT in \eqref{symDCAT_limits2} is calculated using an equivalent expression of the eq.\eqref{normcatRed}.

In the case where all  coordinates $z_i$ tend to 0 ($k=0$), the $\mathbb{0}$-DCAT collapses to a Fock state,
\begin{equation}
	\lim\limits_{\zb\to \bm{0}}|\zb\ra_{\mathbb{0}}^{(N)}=|{\scriptstyle n_0=N,\,n_1=0,\,\ldots,\,n_{D-1}=0}\ra\,,\label{symDCAT_limits3}
\end{equation}
which is the highest weight vector of the $N$-particle symmetric irreducible representation of U$(D)$ that we are considering. This highest weight vector deserves our attention because it is the ground state of the free ($\lambda=0$) LMG U($D$) Hamiltonian (see in Section \ref{LMGsec} for a detailed discussion). The limit \eqref{symDCAT_limits3} has previously been calculated in \cite{DcatDecompositionArxiv} for a general $\mathbbm{c}$-DCAT, giving the so called  \textit{Fock-cat states}. 


\section{Analytical calculation of the $\nu$-moments of the Husimi function of a DSCS}\label{app1}

Here we show in detail the calculations that lead to the expressions of the $\nu$-th moments of the DSCSs (\ref{mumomentD3},\ref{mumomentD3CS}), the DCATs \eqref{mumomentDCAT} and its thermodynamic limit (\ref{mumomentD3},\ref{mumomentDCATlimit},\ref{mumomentDCATlimit2}).

Firstly, the moments of the DSCSs are computed by previously using the highest-weight state $|\psi\ra=|\zb=\bm{0}\ra=(a_0^\dag)^N/\sqrt{N!}|\vec{0}\ra$ (a boson condensate of $N$ atoms in their lower level $i=0$)
according to the equation\eqref{cohD}. Using the scalar product of the DSCSs \eqref{scprod}, we calculate the Husimi function \eqref{HusimiDef} of this state as 
\begin{equation}
	Q_{|\bm{0}\ra}(z)=|\la\zb|\bm{0}\ra|^2=\frac{1}{(1+\zb^\dag\zb)^{N}}\,.
\end{equation}
It is straightforward to perform the integration in the $\nu$-moments formula \eqref{momentsNu} for the Husimi function $Q_{|\bm{0}\ra}(z)$ and arbitrary $\nu$. The integral in $\mathbb{C}^{D-1}$ is mapped to $(\mathbb{R}^+\times[0,2\pi])^{D-1}$ using polar coordinates $z_j=\rho_je^{i\theta_j}$, $d^2z_j=\rho_jd\rho_jd\theta_j$ for all $j=1,\ldots,D-1$. Then, we integrate recursively for all $\rho_j$ from $j=1$ to $j=D-1$, and the equation \eqref{mumomentD3} for $M_{\nu}(|\bm{0}\ra)$ is achieved. The extension  \eqref{mumomentD3CS} from $|\bm{z}=\bm{0}\ra$ to an arbitrary  DSCS $|\bm{z}\ra$ is direct using the $\rmu(D)$ invariance of the Fubini-Study measure $d\mu(\zb)$ in $\mathbb{C}P^{D-1}$.

\section{Thermodynamic limit of the $\nu$-moments of the Husimi function of a $\mathbbm{c}$-DCAT}\label{app2}   


In Eq. \eqref{mumomentDCAT} we have given the  $\nu$-moments of the Husimi function $Q_{|\zb_\mathbbm{c}\ra}$  of a $\mathbbm{c}$-DCAT. This bulky expression acquires a simpler form \eqref{mumomentDCATlimit2} in the thermodynamic limit. Let us prove it.   

We shall initially give some auxiliary results and calculate their Husimi function. First of all, the scalar product of the DSCSs \eqref{scprod} has a Kronecker delta-like thermodynamic limit,
\begin{equation}
	\lim\limits_{N\to\infty}\la \zb'|\zb\ra=
	\begin{cases}
		1 \quad \mathrm{if}\quad  \zb'=\zb\,,\\
		0 \quad \mathrm{if}\quad \zb'\neq \zb\,,
	\end{cases}
\end{equation}
which leads to 
\begin{equation}\label{overlapLimit1}
	\lim\limits_{N\to\infty}\la \zb'|\zb^\mathbbm{b}\ra\la \zb^{\mathbbm{b}'}|\zb'\ra=
	\begin{cases}
		1 \quad\mathrm{if}\quad \zb'=\zb^\mathbbm{b}\text{ and }\zb'=\zb^{\mathbbm{b}'},\\
		0 \quad \text{elsewhere}\,,
	\end{cases}
\end{equation}
as $(1+\zb^\dag\zb^\mathbbm{b})<(1+\zb^\dag\zb)$ for all $\mathbbm{b}\neq\mathbb{0}$ and $\zb$ with non-zero components.
The non-null condition of the last equation implies that $\zb^{\mathbbm{b}}=\zb^{\mathbbm{b}'}$, what leads to $\mathbbm{b}=\mathbbm{b}'$ provided that $z_i\neq0$ for all $ i=1,\ldots,D-1$. Therefore, we begin studying the case where $\zb$ does not have any null component.

The Husimi function of the $\mathbbm{c}$-DCAT \eqref{HusimiDCAT} can also be written using the Husimi function \eqref{HusimiDef} and the $\mathbbm{c}$-DCAT \eqref{DCAT}  definitions,
\begin{align}\label{HusimiDCATAppB}
	Q_{|\zb\ra_{\mathbbm{c}}}(\zb')=&\,|\la \zb'|\zb\ra_{\mathbbm{c}}|^2\\
	=&\, \left(\frac{2^{1-D}}{\mathcal{N}(\zb)_\mathbbm{c}}\right)^2\sum_{\mathbbm{b},\mathbbm{b}'}(-1)^{\mathbbm{c}\cdot(\mathbbm{b}+\mathbbm{b}')}\la\zb'|\zb^\mathbbm{b}\ra\la \zb^{\mathbbm{b}'}|\zb'\ra\nonumber\,.
\end{align}
Since the $\mathbbm{c}$-DCAT normalization $\mathcal{N}(\zb)_\mathbbm{c}$ is non-zero for all $\zb$ (without any null component) and $\mathbbm{c}$, we take the limit of the numerator and denominator of $Q_{|\zb\ra_{\mathbbm{c}}}(z')$ separately. The denominator is, according to the equation \eqref{normcat},
\begin{align}\label{normDCATlimit}
	\lim\limits_{N\to\infty}\mathcal{N}(\zb)_\mathbbm{c}^2=&\,\lim\limits_{N\to\infty}2^{1-D} \frac{ \sum_{\mathbbm{b}} (-1)^{\mathbbm{c}\cdot\mathbbm{b}} (1+\zb^\dag\zb^\mathbbm{b})^N}{(1+\zb^\dag\zb)^N}\nonumber\\
	=&\,2^{1-D} \sum_{\mathbbm{b}}(-1)^{\mathbbm{c}\cdot\mathbbm{b}}\lim\limits_{N\to\infty}\frac{   (1+\zb^\dag\zb^\mathbbm{b})^N}{(1+\zb^\dag\zb)^N}\nonumber\\
		=&\,2^{1-D}\,.
\end{align}
The numerator limit is performed using the equation \eqref{overlapLimit1} and its derived condition $\mathbbm{b}=\mathbbm{b}'$,
\begin{align}
	&\,\lim\limits_{N\to\infty}\sum_{\mathbbm{b},\mathbbm{b}'}(-1)^{\mathbbm{c}\cdot(\mathbbm{b}+\mathbbm{b}')}\la\zb'|\zb^\mathbbm{b}\ra\la \zb^{\mathbbm{b}'}|\zb'\ra\nonumber\\
	=&\,\lim\limits_{N\to\infty}\sum_{\mathbbm{b}}(-1)^{\mathbbm{c}\cdot(\mathbbm{b}+\mathbbm{b})}\la\zb'|\zb^\mathbbm{b}\ra\la \zb^{\mathbbm{b}}|\zb'\ra\nonumber\\
	=&\,\lim\limits_{N\to\infty}\sum_{\mathbbm{b}}Q_{|\zb^\mathbbm{b}\ra}(\zb')\,,
\end{align}
as $(-1)^{\mathbbm{c}\cdot(\mathbbm{b}+\mathbbm{b})}=1$. Therefore, the limit of the $\mathbbm{c}$-DCAT Husimi function is 
\begin{equation}\label{HusimiDCATLimit}
	\lim\limits_{N\to\infty}Q_{|\zb\ra_\mathbbm{c}}(\zb')=2^{1-D}\lim\limits_{N\to\infty}\sum_{\mathbbm{b}}Q_{|\zb^\mathbbm{b}\ra}(\zb')\,.
\end{equation}
The number of humps of $\lim\limits_{N\to\infty}Q_{|\zb\ra_\mathbbm{c}}(\zb')$ in the phase space $\zb'$ will be the number of terms in the sum $\sum_{\mathbbm{b}}$ (right term in eq.\eqref{HusimiDCATLimit}), that is $2^{D-1}$, as showed in the Figures \ref{Figure:HusimiFuncSym3CAT_Variational_LMG} and \ref{Figure:HusimiFunc_c-3CATs_Variational_LMG} for $D=3$ and $\lambda=2.5$.

The next step is calculate the limit of the $\nu$-th power of the Husimi function of the $\mathbbm{c}$-DCAT, $ [Q_{|\zb\ra_\mathbbm{c}}(\zb')]^\nu$ for all $\nu\geq 2$. We split again the limit in numerator and denominator, where the last one is trivial using the same procedure as in \eqref{normDCATlimit}, that is $\lim\limits_{N\to\infty}\mathcal{N}(\zb)_\mathbbm{c}^{2\nu}=(2^{1-D})^\nu$. The numerator limit is 
\begin{align}
	&\,\lim\limits_{N\to\infty}\left(\sum_{\mathbbm{b},\mathbbm{b}'}(-1)^{\mathbbm{c}\cdot(\mathbbm{b}+\mathbbm{b}')}\la\zb'|\zb^\mathbbm{b}\ra\la \zb^{\mathbbm{b}'}|\zb'\ra\right)^\nu\\
	=&\,\lim\limits_{N\to\infty}\sum_{\mathbbm{b}_1,\ldots,\mathbbm{b}_{\nu}}\sum_{\mathbbm{b}'_1,\ldots,\mathbbm{b}'_{\nu}}(-1)^{\mathbbm{c}\cdot\sum_{i=1}^{\nu}(\mathbbm{b}_i+\mathbbm{b}'_i)}\nonumber\\
	&\hspace{30mm}\times\prod_{i=1}^{\nu}\la\zb'|\zb^{\mathbbm{b}_i}\ra\la \zb^{\mathbbm{b}'_i}|\zb'\ra\nonumber\,,
\end{align}
which reduces, with the auxiliary equation \eqref{overlapLimit1}, to
\begin{align}
	&\,\lim\limits_{N\to\infty}\sum_{\mathbbm{b}}(-1)^{\mathbbm{c}\cdot2\nu\mathbbm{b}}\prod_{i=1}^{\nu}\la\zb'|\zb^{\mathbbm{b}}\ra\la \zb^{\mathbbm{b}}|\zb'\ra\nonumber\\
	=&\,\lim\limits_{N\to\infty}\sum_{\mathbbm{b}} [Q_{|\zb^\mathbbm{b}\ra}(\zb')]^\nu\,.
\end{align}
So we have
\begin{equation}\label{HusimiDCATnuLimit}
	\lim\limits_{N\to\infty} [Q_{|\zb\ra_\mathbbm{c}}(\zb')]^\nu=(2^{1-D})^\nu\lim\limits_{N\to\infty}\sum_{\mathbbm{b}} [Q_{|\zb^\mathbbm{b}\ra}(\zb')]^\nu\,.
\end{equation}

Eventually, we can calculate the $\nu$-moments of the $\mathbbm{c}$-DCAT Husimi function, that is
\begin{align}
	\lim\limits_{N\to\infty}M_\nu(|\zb\ra_\mathbbm{c})=\lim\limits_{N\to\infty}\int_{\mathbb{C}^{D-1}} [Q_{|\zb\ra_\mathbbm{c}}(\zb')]^\nu d\mu(\zb')\,.
\end{align}
Employing the equation \eqref{HusimiDCATnuLimit}, and commuting the integral and the limit, the last expression turns into
\begin{align}
	\lim\limits_{N\to\infty}M_\nu(|\zb\ra_\mathbbm{c})=&\,(2^{1-D})^\nu\int\lim\limits_{N\to\infty} \sum_{\mathbbm{b}} [Q_{|\zb^\mathbbm{b}\ra}(\zb')]^\nu d\mu(\zb')\nonumber\\
	=&\,(2^{1-D})^\nu\lim\limits_{N\to\infty} \sum_{\mathbbm{b}}\int [Q_{|\zb^\mathbbm{b}\ra}(\zb')]^\nu d\mu(\zb')\,.
\end{align} 
The new integral is equal to the moment $M_\nu(|\zb^\mathbbm{b}\ra)$ of the DSCS $|\zb^\mathbbm{b}\ra$, which fulfills $M_\nu(|\zb^\mathbbm{b}\ra)=M_\nu(|\zb\ra)=M_\nu(|\bm{0}\ra)$ according to the equation \eqref{mumomentD3CS} and the Fubini-Study measure invariance. In the end, the equation \eqref{mumomentDCATlimit} of the moments of the $\mathbbm{c}$-DCAT in the thermodynamic limit is reached,
\begin{align}
	\lim\limits_{N\to\infty}M_\nu(|\zb\ra_\mathbbm{c})=&\,(2^{1-D})^\nu\lim\limits_{N\to\infty} \sum_{\mathbbm{b}}M_\nu(|\zb\ra)\nonumber\\
	=&\,(2^{D-1})^{1-\nu}\lim\limits_{N\to\infty} M_\nu(|\zb\ra)\,.
\end{align}


When there are only $k$ non-zero components in $\zb$, the even $\mathbb{0}$-DCAT (with $\mathbb{0}\in\mathbb{Z}_2^{D-1}$) reduces to a $\mathbb{0}_K$-DCAT with a smaller parity symmetry $\mathbb{0}_K=[0,\stackrel{(k)}{\ldots},0]\in\mathbb{Z}_2^{k}$ (see the notation of the eq.\eqref{symDCAT_limits2}). Therefore, the equation \eqref{HusimiDCATAppB} turns into
\begin{align}\label{Husimi0DCATAppB}
	&\,\lim\limits_{\zb_L\to\bm{0}_L}Q_{|\zb\ra_{\mathbb{0}}}(\zb')=Q_{|\zb_K\ra_{\mathbb{0}_K}}(\zb')=|\la \zb'|\zb_K\ra_{\mathbb{0}_K}|^2\nonumber\\
	=&\, \left(\frac{2^{-k}}{\mathcal{N}(\zb_K)_{\mathbb{0}_K}}\right)^2\sum_{\mathbbm{b}_K,\mathbbm{b}'_K}(-1)^{\mathbb{0}_K\cdot(\mathbbm{b}_K+\mathbbm{b}'_K)}\la\zb'|\zb_K^{\mathbbm{b}_K}\ra\la \zb_K^{\mathbbm{b}'_K}|\zb'\ra\,,
\end{align}
where $\zb_K=\lim_{\zb_L\to\bm{0}_L}\zb$ (it would be more correct to write it as $(\zb_K,\zb_L=\bm{0}_L)$) and $\mathbb{0}_K,\mathbbm{b}_K,\mathbbm{b}'_K\in\mathbb{Z}_2^{k}$. As previously done in the non-zero case \eqref{normDCATlimit}, the reduced normalization constant of the denominator tends to $\lim_{N\to\infty}\mathcal{N}(\zb_K)_{\mathbb{0}_K}=2^{-k}$, where we have used a generalization of the expression \eqref{normcatRed}. The equation \eqref{overlapLimit1} can be adapted to
\begin{equation}\label{overlapLimit1Red}
	\lim\limits_{N\to\infty}\la\zb'|\zb_K^{\mathbbm{b}_K}\ra\la \zb_K^{\mathbbm{b}'_K}|\zb'\ra=
	\begin{cases}
		1 \quad\mathrm{if}\:\: \zb'=\zb_K^{\mathbbm{b}_K}\text{ and }\zb'=\zb_K^{\mathbbm{b}'_K},\\
		0 \quad \text{elsewhere}\,,
	\end{cases}
\end{equation}
where the non-null value is achieved when $\zb_K^{\mathbbm{b}_K}=\zb_K^{\mathbbm{b}'_K}$, which implies $\mathbbm{b}_K=\mathbbm{b}'_K$. This is true because all the coordinates in $\zb_K$ ($(\zb_K,\zb_L=\bm{0}_L)$ in fact) associated to $\mathbbm{b}_K$ are non-zero by construction. Consequently, the numerator in \eqref{Husimi0DCATAppB} transforms into
\begin{align}
	&\,\lim\limits_{N\to\infty}\sum_{\mathbbm{b}_K,\mathbbm{b}'_K}(-1)^{\mathbb{0}_K\cdot(\mathbbm{b}_K+\mathbbm{b}'_K)}\la\zb'|\zb_K^{\mathbbm{b}_K}\ra\la \zb_K^{\mathbbm{b}'_K}|\zb'\ra\nonumber\\
	=&\,\lim\limits_{N\to\infty}\sum_{\mathbbm{b}_K}\la\zb'|\zb_K^{\mathbbm{b}_K}\ra\la \zb_K^{\mathbbm{b}_K}|\zb'\ra\nonumber\\
	=&\,\lim\limits_{N\to\infty}\sum_{\mathbbm{b}_K}Q_{|\zb_K^{\mathbbm{b}_K}\ra}(\zb')\,,
\end{align}
using in the second line the property \eqref{ChiProp1} of the parity group characters. The thermodynamic limit of the $\tilde{\mathbb{0}}$-DCAT Husimi function is finally
\begin{equation}\label{HusimiDCATLimitRed}
	\lim\limits_{N\to\infty}Q_{|\zb_K\ra_{\mathbb{0}_K}}(\zb')=2^{-k}\lim\limits_{N\to\infty}\sum_{\mathbbm{b}_K}Q_{|\zb_K^{\mathbbm{b}_K}\ra}(\zb')\,.
\end{equation}

From this moment on, it is straightforward to adapt the procedure followed at the beginning for the moments of the $\mathbbm{c}$-DCAT to the $\mathbb{0}_K$-DCAT, arriving to the expression
\begin{align}
	&\,\lim\limits_{N\to\infty}M_\nu(|\zb_K\ra_{\mathbb{0}_K})\\
	=&\,(2^{-k})^\nu\lim\limits_{N\to\infty} \sum_{\mathbbm{b}_K}\int_{\mathbb{C}^{D-1}} [Q_{|\zb_K^{\mathbbm{b}_K}\ra}(\zb')]^\nu d\mu(\zb')\nonumber\\
	=&\,(2^{-k})^\nu\lim\limits_{N\to\infty} \sum_{\mathbbm{b}_K}M_{\nu}(|\zb_K^{\mathbbm{b}_K}\ra)=(2^{k})^{(1-\nu)}\lim\limits_{N\to\infty}M_{\nu}(|\zb\ra)\,,\nonumber
\end{align}
since $M_{\nu}(|\zb_K^{\mathbbm{b}_K}\ra)=M_{\nu}(|\zb\ra)$ \eqref{mumomentD3CS}, and using the parity characters property \eqref{CharPropDelta} for the reduced parity group $\mathbb{Z}_2^k$. The last equation ends the calculations to prove the eq.\eqref{mumomentDCATlimit2} for the $\mathbb{0}$-DCAT. The general case of zero coordinates in the $\mathbbm{c}$-DCAT (see eq.\eqref{mumomentcDCATlimit2}) has been computed with a symbolic calculation software, so the analytical calculations are devoted to future research.

\appendix


\bibliography{bibliografia.bib}

\end{document}